  \documentclass[preprint]{aastex}
 \usepackage{longtable}
 
\newcommand{\Htwo}{H$_2$}            
\newcommand {\HI}     {\ion{H}{1}}      
\newcommand {\SiIII}  {\ion{Si}{3}}      
\newcommand {\NaI}  {\ion{Na}{1}}      
\newcommand {\CaII} {\ion{Ca}{2}}      

\newcommand{\Lya}{Ly$\alpha$}

\newcommand{\FUSE}{{\it FUSE}}
  
\newcommand{\etal} {et~al.}
\newcommand{\kms} {km~s$^{-1}$}

\newcommand{\Ratio} {$N_{\rm H}/E(B-V)$} 

\begin{document}

\title{Variations of Interstellar Gas-to-Dust Ratios \\
 at High Galactic Latitudes } 

\author {J. Michael Shull}
\affil{Department of Astrophysical and Planetary Sciences, CASA \\
University of Colorado, Boulder, CO 80309, USA }  

\author{Georgia V. Panopoulou}
\affil{Department of Space, Earth, \& Environment, \\
Chalmers University of Technology, SE-412~93, Gothenburg, Sweden}

\email{michael.shull@colorado.edu, georgia.panopoulou@chalmers.se}



\begin{abstract}

Interstellar dust at high Galactic latitudes can influence astronomical foreground 
subtraction, produce diffuse scattered light, and soften the ultraviolet spectra 
of quasars. In a sample of 94 sight lines toward quasars at high latitude and low 
extinction, we evaluate the interstellar ``gas-to-dust ratio" \Ratio, using hydrogen
column densities (\HI\ and \Htwo) and far-infrared estimates of dust reddening.  
In the Galactic plane, this ratio is $6.0\pm0.2$ (in units of
$10^{21}~{\rm cm}^{-2}~{\rm mag}^{-1}$).  On average, recent Planck estimates 
of $E(B-V)$ in low reddening sight lines are 12\% higher than those from 
Schlafly \& Finkbeiner (2011), and $N_{\rm HI}$ exhibits significant variations when 
measured at different radio telescopes.  In a sample of 51 quasars with measurements 
of both \HI\ and \Htwo\ and $0.01 \leq E(B-V) \lesssim 0.1$, we find mean ratios 
$10.3\pm0.4$ (gas at all velocities) and $9.2\pm0.3$ (low velocity only) using Planck 
$E(B-V)$ data. High-latitude \Htwo\ fractions are generally small (2--3\% on average),
although 9 of 39 sight lines at $|b| \geq 40\degr$ have $f_{\rm H2}$ of 1--17\%. 
Because FIR-inferred $E(B-V)$ is sensitive to modeled dust temperature $T_d$ and 
emissivity index $\beta$, gas-to-dust ratios have large, asymmetric errors 
at low $E(B-V)$.  The ratios are elevated in sight lines with high-velocity clouds, which 
contribute $N_{\rm H}$ but little reddening.  
In Complex C, the ratio decreases by 40\% when high velocity gas is excluded. 
Decreases in dust content are expected in low-metallicity gas above the Galactic 
plane, resulting from grain destruction in shocks, settling to the disk, and thermal 
sputtering in hot halo gas.  
 
 \vspace{2cm}
 
\end{abstract} 


\section{Introduction}
 
 This paper examines measures of the interstellar ``gas-to-dust ratio", \Ratio, found
 from the total column density of atomic and molecular hydrogen,
 $N_{\rm H} = N_{\rm HI} + 2 N_{\rm H2}$, and selective extinction $E(B-V)$.  In studies 
 with the {\it Copernicus} satellite (Savage \etal\ 1977; Bohlin \etal\ 1978) hydrogen column 
 densities were measured with UV absorption lines (\Lya\ and \Htwo\ Lyman/Werner bands).
 and $E(B-V)$ was obtained from stellar photometry and spectral classification.  Along 75 
 sight lines in the Galactic disk, they found a mean ratio 
 $5.8\times10^{21}~{\rm cm}^{-2}~{\rm mag}^{-1}$, hereafter quoted in standard units of 
 $10^{21}~{\rm cm}^{-2}~{\rm mag}^{-1}$.   A recent survey with the \FUSE\ satellite of 
 sight lines to 129 OB-stars within 5~kpc (Shull \etal\ 2021) found a mean value 
 $6.07\pm1.01$ ($1\sigma$ variance in the distribution) using updated $E(B-V)$ from 
 O-star photometry and spectral types from the Galactic O-star Spectroscopic Survey 
 (Sota \etal\ 2011, 2014).  For a subset of 21 stars at $E(B-V) \leq 0.25$~mag,  the mean 
 ratio was $N_{\rm H}/E(B-V) = 5.83$.  A \FUSE\ survey of 38 translucent sight lines with 
 $A_V \approx$ 0.5--4.7~mag (Rachford \etal\ 2009) found a mean ratio of 5.94, and 
 several large compilations of UV measurements found 6.12 (Gudennavar \etal\ 2012) 
 and 6.2 (Liszt \& Gerin 2023).  We conclude that the interstellar medium (ISM) in the 
 Galactic disk has a consistent mean gas-to-dust ratio of 
 $\sim6 \times 10^{21}~{\rm cm}^{-2}~{\rm mag}^{-1}$ to an accuracy of 3--4\% and 
 with 17\% dispersion.  

However, recent estimates of the gas-to-dust ratio at high Galactic latitudes find 35--45\% 
higher values, using $N_{\rm HI}$ from 21-cm emission and $E(B-V)$ inferred from modeling 
far-infrared (FIR) emission as a tracer of the dust column.   Elevated ratios above the disk plane
might be expected, as dust settles gravitationally to the disk or is blown into the low halo,
where it can be sputtered by hot gas.  Reductions in grain abundance can also result from 
lower gas metallicities and grain disruption in shocks.  In the radio/FIR method, values of $E(B-V)$ 
were taken from two studies (Schlegel \etal\ 1998 and Schlafly \& Finkbeiner 2011) hereafter 
denoted SFD98 and SF11 and available on the IPAC/IRSA reddening 
website\footnote{\url {https://irsa.ipac.caltech.edu/applications/DUST/docs/background.html}}.
In sight lines at latitudes $|b| > 20^{\circ}$ and low reddening, $0.015 < E(B-V) < 0.075$, 
Liszt (2014a,b) found a mean ratio $N_{\rm HI}/E(B-V) = 8.3$, with \HI\ from the
Leiden-Argentina-Bonn (LAB) survey (Kalberla \etal\ 1995) at $36'$ resolution and $E(B-V)$ 
from SFD98.  Lenz \etal\ (2017) found a mean ratio of 8.8, using \HI\ from the HI4PI survey 
(HI4PI Collaboration \etal\ 2016) at $16'$ resolution and $E(B-V)$ from SF11.  They only 
considered atomic gas with $N_{\rm HI} < 4 \times 10^{20}~{\rm cm}^{-2}$ and local standard 
of rest velocities $|V_{\rm LSR}| < 90$~\kms.  Lenz \etal\ (2017) noted that the values obtained 
with the SFD98 maps are somewhat higher than the 8.3 value obtained by Liszt (2014b), 
who did not use the 12\% SF11 correction to the SFD98 calibration.  Liszt \& Gerin (2023) 
quoted a mean high-latitude ratio of 8.3.

The question arises whether the elevated gas-to-dust-ratios at high latitude are real or the result 
of different methods and calibrations.  To investigate the reliability of the gas-to-dust ratio \Ratio\ 
we examine measurements of both the numerator and denominator, considering their systematic 
uncertainties. Both Liszt (2014a,b) and Lenz \etal\ (2017) used only \HI, noting that corrections for 
\Htwo\ are normally small for $E(B-V) < 0.08$.   This may not always be the case, as seen in our 
sample of quasar sight lines with both \HI\ and \Htwo\ data.  Errors in measurements of $N_{\rm HI}$ 
can result from telescope beam sizes, stray radiation removal, baseline determination and calibration, 
and methods of integrating the 21-cm emission profile over velocity.  Following studies by 
Wakker \etal\ (2011), we have compared $N_{\rm HI}$ in surveys with radio telescopes whose beam 
sizes range from $36'$ to $9'$, including those from the Green Bank Telescope and the HI4PI survey.   
We also assess the accuracy of FIR thermal emission inferences of equivalent optical extinction.  
Several reddening maps have been presented in the literature, exhibiting systematic differences 
(e.g., Lenz \etal\ 2017; Chiang \& M\'enard 2019).   For example, the reddening map presented in 
Planck Collaboration XI (2014) has larger $E(B-V)$ at high Galactic latitude compared to SF11 
(Lenz \etal\ 2017; Casandjian \etal\ 2022).   All of these uncertainties result in large and asymmetric 
errors in the gas-to-dust ratio, especially at low reddening.

Our data set includes sight lines to 94 quasars at high Galactic latitude, most at $|b| > 25\degr$. 
Indeed, 78 lie at $|b| \geq 40\degr$.   Figure 1 shows a polar projection of the quasar locations 
and ratios between \HI\ column density and two FIR-inferred $E(B-V)$ maps (see Section 2.3).  
The \HI\ column density has been computed within the same velocity range [-90, 90] \kms\ 
used to derive the gas-to-dust ratio in Lenz \etal\ (2017).  The spatial variation of the ratio over 
the sky differs between the two reddening maps, indicating systematic reddening uncertainties,
probably arising from FIR modeling of dust temperature and emissivity spectral index.

In Section 2 we discuss the sources of data used to evaluate the gas-to-dust ratio, including 
column densities $N_{\rm HI}$, $N_{\rm H2}$, selective extinction $E(B-V)$, and their 
uncertainties. We examine the measurements used to construct \Ratio\ ratios along the sight 
lines to 94 high-latitude quasars.  In particular, we explore uncertainties in 21-cm measurements 
of $N_{\rm HI}$ from different radio telescopes and uncertainties in inferring $E(B-V)$ from FIR 
emission and grain emissivity models.  In Section 3 we present data from our survey, which 
confirm previous observations of elevated gas-to-dust ratios at $|b| > 30\degr$.  However, we 
suggest large uncertainties in the ratio.  An important effect in elevated ratios is the reduced 
dust content in Galactic high velocity clouds (HVCs) and some intermediate velocity clouds (IVCs).  
In Section 4, we summarize our results and their implications for astronomical foreground 
subtraction, diffuse scattered light, and reddening corrections for the ultraviolet spectra of 
quasars.  We stress the importance of identifying the systematic uncertainties in $\log N_{\rm H}$ 
and $E(B-V)$, which are almost certainly much larger than the small values ($\leq5$\%) 
commonly listed in observational tables.  

\section{Data Compilation for Gas-to-Dust Ratios} 

Our full sample includes 94 quasars at high Galactic latitude (Table 1).  The first 55 quasars 
listed (group 1) have both \HI\ and \Htwo\ column densities, a primary sample that includes 
47 sight lines with IVCs and 18 with HVCs.   Nine AGN lie behind Complex~C (Figure~2),
an extended structure of high-velocity gas with metallicities 10--30\% solar (Wakker \etal\ 1999; 
Collins \etal\ 2007; Shull \etal\ 2011).  We also analyzed 39 additional quasars (group 2 in Table 1) 
in which only \HI\ column densities were available.  In contrast to UV studies of \HI\ and \Htwo\ in 
the Galactic disk, the high-latitude measurements of the dust-to-gas ratio (Liszt 2014a,b; 
Lenz \etal\ 2017) used 21-cm emission for $N_{\rm HI}$ and FIR emission as a proxy for $E(B-V)$.  
As described in Appendix~A, these techniques can introduce systematic errors in the gas-to-dust 
ratio, particularly when $E(B-V) \lesssim 0.04$ mag.  Toward high-latitude quasars, reddening 
uncertainties often dominate the propagated errors in the ratio \Ratio.   However, we also found 
cases in which measurements of $N_{\rm HI}$ differ by 10--30\% among different telescopes
with a range of beam sizes.  

We use two reddening maps based on previous modeling of the thermal dust emission: 
(1) recalibration of the SFD98 map by SF11; (2) the map presented in Planck 
Collaboration XLVIII (2016).  The 2016 Planck map employed the GNILC technique 
(Generalized Needlet Internal Linear Combination) which uses spatial information from 
angular power spectra and diffuse component separation to reduce contamination by 
cosmic infrared background radiation.  These values are preferred over those from
Planck-DL (Planck Collaboration XXIX 2016).  Values of $E(B-V)$ for both SFD98 and SF11 
are available on the IPAC/IRSA reddening website.  Although that website quotes both values, 
we only list SF11 values, which are 86\% of SFD98 owing to re-calibration of colors using 
stars from the Sloan Digital Sky Survey\footnote{The re-calibrated SF11 values of $E(B-V)$ 
are 86\% those of SFD98, but were not applied by Liszt (2014b). The 0.86 factor is less  
than the conversion factor described in Section 2.2.1 of Lenz \etal\ (2017) where 0.884 
is for the Landolt-V filter, and 0.865 is for the CTIO-V filter.  Lenz \etal\ (2017) also noted
a mean ratio $N_{\rm HI}/E(B-V) = 8.2$, using pixels in a revised reddening map 
(Schlafly \etal\ 2014) based on Pan-STARRS1 optical photometry of 500 million stars.}.   
The SF11 study found that the SFD98 map needed 
a re-calibration by a factor 0.86 (0.865 in their Table 6 for the CTIO-V filter). The map has a 
FWHM of $6.1'$. We query this map using the IPAC/IRSA website, which provides $1\sigma$ 
variances averaged over neighboring $5'$ pixels around each line of sight.   We also downloaded 
the original SFD98 map projected into a HEALPix grid of $N_{\rm side} = 2048$ and applied the 
re-calibration correction 
factor\footnote{The map is provided on \url{lambda.gsfc.nasa.gov} by Chiang (2023).}.  

We investigated shifts in the gas-to-dust ratio when we use $E(B-V)$ from Planck-GN 
(Planck Collaboration XLVIII 2016) instead of SF11.   For the 48 AGN sight lines with 
$E(B-V) \leq 0.02$, the GN reddenings are 15\% higher than SF11 on average 
(15 are lower, 33 are higher).  For the 81 AGN with $E(B-V) \leq 0.04$, the GN reddenings 
are 12\% higher on average.  Thus, using GN reddening instead of SF11 would reduce 
the gas-to-dust ratios by 11--13\%.   For the 11 sight lines with $E(B-V) < 0.01$, the ratio 
reductions are even larger.  In Table~1, we highlight these uncertain sight lines in boldface 
(four in group~1, seven in group~2). 

\subsection{Atomic Hydrogen ($N_{\rm HI}$) } 

The \HI\ column densities came from 21-cm spectra in several surveys.  All 94 AGN appear
in the \HI\ compilation of Wakker \etal\ (2003), who published velocity profiles from a variety
of radio telescopes, primarily the Green Bank 140-ft (GB), 100-m Green Bank Telescope (GBT), 
Leiden-Dwingeloo Survey, Effelsberg, and Villa Elisa.  This 2003 paper has been used in 
many studies of interstellar gas (metallicities, \Htwo, and ionized phases) because it contains 
velocity-component fits to the emission profiles.  This allows  to keep separate accounts
of low-velocity \HI\ and HVCs and IVCs, when present.  However, Gaussian fitting can 
sometimes give spurious total column densities.  
 
 We also study these sight lines using the HI4PI all-sky survey, which combined data 
from the Effelsberg and Parkes radio telescopes with an angular resolution of $16'$ 
(HI4PI Collaboration \etal\ 2016). We use two related data products: 
(a) the $N_{\rm HI}$ map computed by integrating the \HI\ spectra over the entire velocity 
range of $|V_{\rm LSR}| < 600$~\kms; and  
(b) the $N_{\rm HI}$ map obtained by Lenz \etal\ (2017) who integrated over low-velocity 
gas, $|V_{\rm LSR}|  < 90$~\kms\ excluding HVC emission, and masked out regions with 
$N_{\rm HI} > 4 \times 10^{20}~{\rm cm}^{-2}$. We refer to the former as NHI--HI4PI(all) 
and the latter as NHI--HI4PI(90). Both maps are provided in HEALPix with 
$N_{\rm side} = 1024$.

Liszt (2014a,b) used \HI\ column densities from the LAB survey ($36'$ beam), and
Lenz \etal\ (2017) used the HI4PI survey ($16.1'$).  For the \HI\ column densities and
21-cm spectra presented by Wakker \etal\ (2003) the beams range from $35'$ in the 
Leiden-Dwingeloo Survey (Hartmann \& Burton 1997) to $16.1'$ with the Parkes Telescope 
and 9--10$'$ (GBT and Effelsberg).  A large beam could include small-scale \HI, and 
velocity component fitting could be more difficult.  These should not be major issues at high 
latitudes except in cases of small-scale structure.  Nevertheless, there is a potential mismatch 
to the resolutions of the space-borne telescopes:  IRAS (4--5$'$ at 100~$\mu$m) and 
Planck (5$'$ at 350--857~$\mu$m). With its $9.1'$ beam and avoidance of stray radiation
due to its off-axis mount, the GBT should provide reliable column densities.  This was the 
conclusion in a previous comparison (Wakker \etal\ 2011) who also identified 10\% offset 
in $N_{\rm HI}$  from data taken in the LAB survey, owing to a ``spurious broad component" 
with $N_{\rm HI} \approx 5\times10^{19}~{\rm cm}^{-2}$.  

In Table 2 we compare $N_{\rm HI}$ measured by radio telescopes with various beam 
sizes toward 36 AGN in common with those tabulated by Wakker \etal\ (2011).  The column 
densities in Table 1 (from Wakker \etal\ 2003) are higher on average by $+0.059$ (dex) 
in $\log N_{\rm HI}$ relative to those from GBT.  Columns from the GB 140-ft are higher by 
$+0.017$ (dex), and those from the HI4PI survey are higher by $+0.023$ (dex).   Quoted 
measurement errors in $\log N_{\rm HI}$ are typically $\pm$ 0.01--0.03.  We consider 
differences $\Delta \log N_{\rm HI} \geq 0.050$ to be discrepant and highlight them in boldface.  
Figure~3 shows the differences in $N_{\rm HI}$  between HI4PI and 
Wakker \etal\ (2003).  Although the average offset differences are comparable to the quoted 
errors on $\log N_{\rm HI}$, several AGN sight lines (e.g., 3C~273, Mrk~279, Mrk 335, 
Mrk~1383, PG~1259+593, PG~0953+414, Ton~S180) exhibit somewhat larger offsets
(0.05--0.12 dex).  These include 5 of 32 sight lines measured by both GBT and HI4PI.

Systematic errors in $N_{\rm HI}$ could arise from calibration, stray-radiation contamination, 
or small-scale structure influenced by beam size.  Integration of antenna temperature over
 the 21-cm velocity profile could also result in variations in the total column density. This may 
 be the case in complex 21-cm profiles such as Mrk~279 and PG~1259+593.
 Reducing the \HI\ column densities by 12--15\% would lower the gas-to-dust 
ratios, as would increasing $E(B-V)$ in low reddening sight lines.  

\subsection{Molecular Hydrogen ($N_{\rm H2}$) }  

Of the 94 quasars in our survey, 55 quasars (group 1) have both \HI\ and \Htwo\ column
densities.  The \Htwo\ column densities were measured by FUSE far-ultraviolet (FUV) 
absorption-line spectra.  With some survey overlap, these came from 45 sight lines in
Gillmon \etal\ (2006), 18 in Wakker (2006), and one each from Richter \etal\ (2001b), 
Collins \etal\ (2003), and Fox \etal\ (2005).  Notes on overlapping \Htwo\ data from these
surveys and corrections for specific sight lines are provided in Appendix B.  In most 
of the 55 sight lines, the \Htwo\ column densities are much smaller than those of \HI.  
In 30 of the 39 AGN sight lines at latitudes $|b| \geq 40\degr$, the molecular fraction 
$f_{\rm H2} \equiv 2N_{\rm H2}/N_{\rm H}$ ranges between $3 \times 10^{-6}$ and 
$5 \times 10^{-4}$, often making \Htwo\ a negligible contributor to the total $N_{\rm H}$ 
in sight lines with column densities below the atomic-to-molecular transition at
$\log N_{\rm H} \approx 20.38\pm0.13$ seen at high latitude (Gillmon \etal\ 2006).  
However, nine sight lines at $|b| \geq 40\degr$ have $f_{\rm H2} > 0.01$ 
(range 1--17\%) and are listed in Table 3, together with $\log N_{\rm H}$ and 
$E(B-V)$.   These 9 sight lines have $E(B-V) =$ 0.02--0.06 (SF11 scale), which is below
the observed \HI-to-\Htwo\ transition at $E(B-V) \approx$ 0.08--0.10  seen in low-latitude 
surveys (Bohlin \etal\ 1978;  Shull \etal\ 2021).  Shifts in the transition to lower $N_{\rm H}$  
toward high-latitude quasars are influenced by lower gas metallicities, lower dust content, 
and reduced far-UV (\Htwo-dissociating) radiation fields; see Browning \etal\ (2003) and 
Gillmon \etal\ (2006) for models.

We also analyzed 39 additional quasars (group 2 in Table 1) in which only \HI\ column 
densities were available.  Because these AGN are all at $|b| \geq 40\degr$ with 
$E(B-V) < 0.056$, the \Htwo\ contributions could be small in most cases.  However, 
17 of the 39 sight lines have column densities $\log N_{\rm H} \approx$ 20.25--20.67,
near or above the \HI-to-\Htwo\ transition.  Thus, some of the group-2 sample would likely
require corrections (1\% to 10\%) for undetected \Htwo.  In our Tables, we separate the 
two groups (55 and 39 AGN) and conduct independent statistical analyses.

\subsection{Reddening Maps }  
 
The dust optical depth map presented in Planck Collaboration XLVIII (2016) was obtained by 
modeling the Galactic thermal dust emission and separating contributions from the Cosmic 
Infrared Background (CIB). The GNILC method was applied to the Planck 2015 data 
and the IRAS $100~\mu$m  map. The final optical depth map has a $5'$ beam size and is 
provided at HEALPix $N_{\rm side} =  2048$ resolution. We multiply the optical depth map at 
353 GHz by the conversion factor $1.49 \times 10^4$ to obtain $E(B-V)$ in magnitudes.
Henceforth, we will refer to these techniques as Planck-GN, and the reddening map as 
the GN map.  When comparing to $N_{\rm HI}$ maps from HI4PI and Lenz \etal\ (2017) we 
smooth the reddening maps to $16.1'$ resolution to match the \HI\ data.  As discussed in 
Appendix E of Planck Collaboration XI (2014), the choice of filter used to compute $E(B-V)$ 
affects the conversion from dust emission to reddening. The Planck team used the filter 
transmission of the Johnson photometric system (M.-A. Miville-Desch\^enes, private 
communication).  The recalibration factor of SFD98 in this system is 0.884, as used in 
Lenz \etal\ (2017). 

Toward the AGN at high Galactic latitude and low extinction, the inferred values of $E(B-V)$ 
have large uncertainties, and may in fact be underestimated. For the quasars in our sample, 
the mean (FIR-inferred) values from SFD98 and SF11 are low:  $E(B-V) = 0.030$ for group~1 
(55 quasars) and 0.019 for group~2 (39 quasars).  Figure 4 compares differences 
in $E(B-V)$ between Planck-GN and SF11 values.  On average, the Planck-GN values are 
12\% larger than those from SF11 towards the AGN in our sample, including those for the 
nine Complex C sight lines (plotted in red). Because of their improvement over previous 
Planck products, we use the GN maps, which were recommended for thermal dust science.  
The visual extinction is often estimated as $A_V = R_V E(B-V)$, with a commonly adopted 
value of $R_V = 3.1$.  This adds further systematic uncertainty, as Peek \& Schiminovich 
(2013) found that no single value of $R_V$ is valid over the entire high-latitude sky.   

Using FIR emission from foreground dust to estimate the equivalent optical reddening requires 
sophisticated models of the grain temperature and emissivity (e.g., Draine \& Li 2007; 
Compi\`egne \etal\ 2011; Hensley \& Draine 2021) which depend on grain composition, size 
distribution, and solid-state properties.  In Appendix C we discuss the dependence of FIR 
(353 GHz) optical depth and dust radiance on the dust temperature ($T_d$) and emissivity 
index ($\beta$), where emissivity $\epsilon(\nu) \propto \nu^{\beta}$.  Combining the scaling 
of dust radiance ${\cal R} \propto \tau_{353} T_d^{4+\beta} \propto T_d^{3 +\beta}$ with the 
observed anti-correlation ($\beta \propto T_d^{-\alpha}$), we find that radiance is sensitive to 
small changes in the two parameters ($\alpha$, $\beta$),
 \begin{equation} 
     \frac {{\cal R}} {{\cal R}_0} = \left( \frac {\beta} {\beta_0} \right)^{-(3 + \beta)/\alpha}   \;  .
\end{equation} 

Table 4 shows the dust parameters adopted in several FIR papers. Over the range of 
indices, $\beta = 1.6 \pm 0.2$, adopted in the 2016 Planck-GNILC study and with 
$\alpha = 2/3$, the radiance factor would change by a factor of 2.3 about ${\cal R}_0$.
Table 5 lists the 100~$\mu$m surface brightness $I(100~\mu$m), FIR-modeled dust 
temperature $T_d$, and inferred color excess, $E(B-V)$ from SF11. The listed error bars 
are $1\sigma$ variances in $E(B-V)$ over neighboring $5'$ pixels.  The true uncertainties 
are likely much larger.  Column~4 lists  $I(100~\mu$m)/$E(B-V)$, with errors propagated 
from relative errors on $I(100~\mu$m) and $E(B-V)$ added in quadrature.  
Several AGN sight lines have large relative errors on $E(B-V)$ and $I(100~\mu$m), with 
$I(100~\mu$m)/$E(B-V)$ uncertain by 10-15\%.  The surface brightnesses typically 
range from 0.4--2~MJy~sr$^{-1}$, with mean values 1.61~MJy~sr$^{-1}$ (first group of 55) 
and 1.07~MJy~sr$^{-1}$ (second group of 39).  The mean dust temperatures are 
$\langle T_d \rangle = 17.900~{\rm K} \; (17.912, 17.892)$ and the mean ratios are
$\langle I(100~\mu{\rm m})/E(B-V) \rangle = 56.22~{\rm MJy~sr}^{-1}~{\rm mag}^{-1} \; 
(56.24, 56.20)$.  Here, the first values are for all 94~quasars, and the numbers in 
parentheses denote the means for groups 1 and 2.   The uniformity in mean dust 
temperatures is surprising, with $1\sigma$ variance $\sigma(T_d) = 0.213~{\rm K}$ 
(1.2\% in group 1). The correspondence between FIR flux and reddening is good, with 
10\% variance $\sigma(I/E) = 5.63~{\rm MJy~sr}^{-1}~{\rm mag}^{-1}$ in the distribution 
of $I(100~\mu{\rm m})/E(B-V)$.   However, because of the sensitivity of the modeled dust 
column density to $T_d$ and $\beta$, small changes can alter the inferred $E(B-V)$.  

Given the uncertainties in the dust modeling, the systematic errors on $E(B-V)$ are likely 
much larger than those quoted in the IPAC/IRSA tables ($1\sigma$ variances).  For these 
reasons, we are suspicious of the accuracy of the ratios \Ratio\ for AGN sight lines with 
$E(B-V) \lesssim 0.04$.  In the next section, we examine these issues statistically for 
various sub-samples.

\section{Survey Results} 

Table 1 presented the gas column densities, inferred $E(B-V)$, and corresponding ratio \Ratio.  
All 94 quasars are lightly reddened, with $E(B-V)$ extending from 0.005 to 0.110 (SF11 scale).  
The first group of 55 quasars with both \HI\ and \Htwo\ column densities has mean 
$E(B-V) = 0.030$, while the second group (\HI\ only) has mean $E(B-V) = 0.019$.  
The difference likely arises from the somewhat higher latitudes in the second group.  
In group 1, we combine molecular hydrogen column densities $N_{\rm H2}$ with those 
of atomic hydrogen $N_{\rm HI}$ to arrive at the total hydrogen column density 
$N_{\rm H} = N_{\rm HI} + 2 N_{\rm H2}$.  Columns 9 and 10 in Table 1 list 
dust-to-gas ratios for gas at all velocities and for low-velocity gas only.

In Section 3.1 we discuss the statistical changes of excluding these sight lines in modified 
sub-samples (51 in group 1 and 32  in group 2).   
In Section 3.2, we compare the gas-to-dust ratios in sight lines that contain HVCs and IVCs.  
In Section 3.3 we discuss Planck all-sky maps of dust extinction and differences from 
earlier studies.  

\subsection{Gas-to-Dust Ratios for QSO Sub-samples}

Table 6 shows mean values of the ratios \Ratio\ for sub-samples of the 94 high-latitude 
QSOs.  The first group includes the 55 quasars for which we have both \HI\ and \Htwo\ column 
densities.  The second group includes 39 quasars with only \HI.  In each group, we list two 
mean ratios:  one for all \HI\ velocity components fitted by Wakker \etal\ (2003) and a second 
for low-velocity gas with $|V_{\rm LSR}| < 90$~\kms.  Elevated ratios are evidence that 
dust grains are deficient in high-velocity gas. For the 55 quasars, the mean ratio using
Planck-GN values of $E(B-V)$ is 11\% higher than SF11. For the 39 quasars in group~2, the 
mean ratio for Planck-GN is 12\% higher. 
In the low-velocity statistics we excluded all HVCs and most of the IVCs, except for 17 IVCs 
in well-known structures containing \Htwo\ and included with low-velocity gas. The two modified 
sub-samples in Table 6 omit the 11 sight lines with highly uncertain $E(B-V) \leq 0.01$.  The 
mean ratios of the  sub-samples and excluded sight lines are listed for comparison.  

Figure 5 plots the 55 (group 1) individual gas-to-dust ratios vs.\ $E(B-V)$.  Some of the 
elevated ratios are unreliable because of large reddening uncertainties at $E(B-V) \leq 0.04$.   
The two values of reddening in the IPAC/IRSA tables (SFD98 and SF11) differ by 14\% because 
of recalibration.  Systematic differences in $E(B-V)$ can arise because of FIR modeling 
sensitivity to dust parameters ($T_d$, $\beta$) as discussed in Appendix~C.  
Gas-to-dust ratios appear high in HVCs because of low grain content, which adds \HI\ 
column density to the sight line without dust reddening.  It remains unclear whether the
dust deficiency arises from low metallicity or grain disruption in shocks (or both).  Low dust 
content in HVCs and IVCs was also noted in the Planck papers, for example Section 6.3 of 
Planck Collaboration XXIV (2011), and in Figures 4 and 5 in Lenz \etal\ (2017) that illustrate 
the dependence of $N_{\rm HI}/E(B-V)$  on \HI\ velocity.  Their range of ratios is even larger 
than found here, probably due to different sky selections.  

Complex~C is a good example of HVC effects, demonstrating the importance of
keeping separate account of the high velocity gas.  Table~7 shows the gas-to-dust 
ratios for nine quasar sight lines passing through this gaseous structure.  The last 
four columns list the ratios derived for gas at all velocities and then omitting \HI\ in 
the HVCs.  The ratios are shown for $E(B-V)$ taken from both SF11 and Planck-GN.  
In both cases, the mean gas-to-dust ratios drop 40\% when one excludes HVCs.  
In Complex~C, gas velocity is a major factor in the elevated ratios.    

Table~7 also illustrates differences in $E(B-V)$ estimates from SF11 and Planck-GN 
among the Complex~C sight lines.  On average, the Planck-GN estimate is 23\% higher
than SF11, over a range in SF11 reddening from $E(B-V) = 0.0053$ to 0.0344.  The 
three sight lines with $E(B-V) < 0.010$ are likely quite uncertain, resulting in large, 
asymmetric errors on $N_{\rm H}/E(B-V)$.  
For example, PG~1626+554 has $\log N_{\rm H} = 20.053$ (all velocities) and 19.936 
(low-velocity gas) but with different values of $E(B-V) = 0.0053$ (SF11) and 0.0137 (GN).  
Adopting the GN reddening instead of SF11 reduces $N_{\rm H}/E(B-V)$ from 21.3 to 
8.24 (all-velocities) and from 16.3 to 6.28 (low-velocity gas).  Similarly, toward Mrk~817, 
where $\log N_{\rm H} = 20.085$ (all velocities), $E(B-V) = 0.0059$ (SF11), and 0.0115 
(GN), the gas-to-dust ratio drops by a factor of two, using GN instead of SF11.
 
\subsection{Distinguishing Low-Velocity and High-Velocity Gas}

Here we examine the possibility of different grain abundances in high velocity gas.  
Specifically, we explore the dust-to-gas ratios after excluding HVCs and some IVCs.  
To distinguish low-velocity gas from higher velocity clouds, we tabulated the column densities 
of the velocity components and performed statistics with and without HVC/IVC gas.  This 
allowed us to assess whether some of the gas is deficient in dust as a result of grain processing 
in interstellar shocks (Draine \& Salpeter 1979; Seab \& Shull 1983; Jones \etal\ 1996; 
Slavin \etal\ 2004).  The \HI\ column densities in Table 1 (columns 5 and 6) were determined 
by summing the Gaussian component fits (Wakker \etal\ 2003) in the 21-cm spectra.  

Most of the HVCs  (Wakker \& van Woerden 1997) with velocities $|V_{\rm LSR}| \geq 90$~\kms\ 
in the local standard of rest show no evidence for dust, probably because of reduced metallicity 
or shock destruction of grains. In some sight lines, HVCs provide a sizable portion of the 21-cm 
emission (Wakker \etal\ 2003; Collins \etal\ 2007; Shull \etal\ 2011; Martin \etal\ 2015; 
Panopoulou \& Lenz 2020).  IVCs have broadly been classified (Albert \& Danly 2004) as having 
$|V_{\rm LSR}|$ between 20--90~\kms.  In recent surveys, the IVC ranges were chosen as 
30--90~\kms\ (Richter \etal\ 2003) and 40--90~\kms\ (Lehner \etal\ 2022). Located in the lower 
Galactic halo, IVCs display a variety of physical conditions, with gas metallicities near solar values 
(Wakker 2001; Richter \etal\ 2001a, 2003), but refractory element abundances that suggest 
some grain disruption.  

From the 21-cm spectra of our sample, we grouped the \HI\ emission components into three 
velocity categories:  
HVCs ($|V_{\rm LSR}| \geq 90$~\kms), IVCs ($|V_{\rm LSR}| =$ 30--90~\kms), and 
low-velocity gas. In the first group of 55 AGN sight lines with both \HI\ and \Htwo\ data, 
we identified HVCs in 18 quasar sight lines  (33\% coverage) and IVCs in 39 sight lines
(73\% coverage).  In the second group of 39 quasars with only \HI\ data, we identified 
HVCs toward 3 quasars (8\%) and IVCs toward 27 quasars (69\%).  The difference in HVC
incidence may be an effect of the higher Galactic latitudes of the AGN in group~2.  
Nine of the quasars in group~1 were targeted to study HVC Complex~C.  In addition,
high-latitude absorbers may be more ionized, with spatial extents greater than those seen
in \HI.   Previous UV studies of IVC/HVC  ionized gas in the strong \SiIII\ 1206.500~\AA\
absorption line found large sky-covering fractions, $f_c = 0.81\pm0.05$ (Shull \etal\ 2009) 
and $f_c = 0.77\pm0.06$ (Richter \etal\ 2017).  

FIR emission has been observed in some IVCs (e.g., Planck Collaboration XXIV 2011; 
Planck Collaboration XI 2014), and the infrared cirrus was shown to correlate with \Htwo\ 
absorption (Gillmon \& Shull 2006). In our statistical analysis of velocity effects, we included 
17 strong IVCs with the low-velocity gas. These sight lines are marked by asterisks in 
column~10 of Table~1.  We excluded all HVCs and most IVCs from the column densities 
of low velocity gas.  The excluded IVCs have velocities well separated from the low-velocity 
21-cm emission near the LSR.  Included with the low-velocity gas were 10 of the 39 IVCs in 
group~1, and 7 of the 27 IVCs in group~2.  They are all well-known structures: eight 
sight lines through the Intermediate Velocity Arch (IV Arch), three through the S1 cloud, 
four through IV18, and one each through IV19 and IV26.   

Planck Collaboration XXIV  (2011) noted that IVCs had different FIR properties, with different 
emission cross sections, often 50\% lower compared to the low-velocity clouds.  
There is also evidence for grain disruption in intermediate velocity absorbers, including 
the ``Routly-Spitzer effect" (Routly \& Spitzer 1952) in which elevated \CaII/\NaI\ ratios 
are observed at increasing cloud velocity.  Similar effects are observed in the rising 
abundances of refractory elements (Si, Fe) with increasing cloud velocity (Shull \etal\ 1977).  
There has been no strong evidence for dust emission in HVCs (e.g., Wakker \& Boulanger 
1986;  D\'esert \etal\ 1988) other than an unconfirmed claim of IR emission in one HVC  
(Miville-Desch\^enes \etal\  2005).   Fox \etal\ (2023) reported indirect evidence for some
dust in Complex C, based on sub-solar differential abundance ratios of refractory elements 
(Fe/S, Si/S,  Al/S) relative to sulfur, which is assumed to be undepleted.  Similar depletion 
measurements have been seen in the Leading Arm of the Magellanic Stream 
(Richter \etal\ 2018). 
  
\subsection{All-Sky Maps}

The sight lines through Complex~C suggest that both low dust content and uncertainties in 
FIR estimates of $E(B-V)$ could be responsible for some of the elevated ratios along high
latitude sight lines with low reddening.   We have compared the SF11 estimates of $E(B-V)$ 
to values from Planck Collaboration XLVIII (2016) denoted here as Planck-GN.  
Polar projection maps in Figure~6 illustrate the differences, which often track changes 
in the gas-to-dust ratio in the maps of Figure~1.  Features in the Northern Galactic hemisphere 
that appear yellow in the GN ratio map correlate with locations of IVCs, where both SFD98 
and SF11 overestimate the total reddening compared to Planck-GN.   The bright feature in 
red near the North Galactic Pole at ($\ell, b) = (260-330\degr,  80-84\degr)$ is Markkanen's 
cloud (Markkanen 1979).  This feature is also known as the North Galactic Pole Rift, seen 
in \HI\ (Puspitarini \& Lallement 2012) and appearing as a foreground shadow in X-rays 
(Snowden \etal\ 2015).  This region is known (Planck Collaboration XI 2014) to have a low 
dust emission spectral index ($\beta$) compared to the rest of the high-latitude sky.   
It shows up in the reddening difference map because SFD98 assumed a constant $\beta$, 
whereas Planck-GN fit for the emission index.  

Figure 7 shows three distributions of gas-to-dust ratios, using different values of
$E(B-V)$.  The three colored curves show $N_{\rm HI}/E(B-V)$ for low-velocity \HI, 
with $E(B-V)$ taken from the SF11 re-calibration (blue), SFD recalibrated with 0.884 
(orange), and Planck-GN (green). We find mean ratios of 9.3 (SF11) and 8.6 (Planck) 
in units of $10^{21}~{\rm cm}^{-2}~{\rm mag}^{-1}$.  The two vertical (dotted, dashed) 
lines show the mean high-latitude ratios (8.8 and 8.2) quoted in Lenz \etal\ (2017).  
The two vertical solid lines show the Galactic disk-plane values (5.8 and 6.07) from 
Bohlin \etal\ (1978) and Shull \etal\ (2021).

\section{Summary and Conclusions }

The goal of our study was to assess the accuracy and reliability of measurements of 
the gas-to-dust ratio \Ratio\ toward high-latitude extragalactic sources. Using radio/FIR 
techniques, past studies (Liszt 2014a,b; Lenz \etal\ 2017; Liszt \& Gerin 2023) found 
35--45\% higher ratios than established values in the Galactic disk plane.  There are
many astrophysical processes that could segregate dust from gas  
(Hensley \& Draine 2021; Shull \etal\ 2021) to produce a deficit of interstellar dust grains 
above the disk plane.  Dust could settle to the disk, be radiatively elevated into the halo, 
or be transported by supernova-driven outflows.  Most HVCs exhibit little or no evidence 
for dust, either because of low metallicity or shock destruction.  Dust elevated above the
Galactic plane will come into contact with hot gas, with grain sputtering lifetimes of  
$t_{\rm sp} \approx (1~{\rm Gyr})(10^{-3}~{\rm cm}^{-3}/n_e)$ at $10^{6.0-6.5}$~K.

Observations of gas and dust at high and low Galactic latitudes employ different methods 
and calibrations.  In the Galactic disk surveys, the hydrogen (\HI, \Htwo) column densities 
were measured from UV absorption toward OB-type stars, with $E(B-V)$ inferred from 
stellar photometry and intrinsic colors assigned to spectral classification.  Most high-latitude 
gas measurements employ \HI\ 21-cm emission toward extragalactic targets, and $E(B-V)$ 
is inferred from models that convert FIR dust emission to the corresponding optical extinction.  
In some cases, \Htwo\ measurements are available toward AGN, but often not.  

From our survey of 94 AGN, we confirm previous observations of elevated gas-to-dust ratios 
at high Galactic latitude (Liszt 2014a,b; Lenz \etal\ 2017). However, we found systematic 
uncertainties in measurements of both the numerator $N_{\rm HI}$ and denominator $E(B-V)$ 
of the ratio.  The different ratios found with the two techniques are seen primarily at high latitudes 
and in sight lines with $E(B-V) \leq 0.04$.   From sub-samples of the 94 AGN, examining the 
measurements $N_{\rm HI}$, $N_{\rm H2}$, and $E(B-V)$, we came to several conclusions 
about offsets and uncertainties:
\begin{itemize}

\item Values of $E(B-V)$ from Planck-GN generally exceed those from SF11 towards the 
   AGN in our sample.  On average, we found their ratio (GN/SF11) to be 15\% higher for 
   48 AGN with $E(B-V) \leq 0.02$ and 12\% higher for 81 AGN with $E(B-V) \leq 0.04$.  
   
\item  Measurements with the GBT 100~m telescope exhibit $N_{\rm HI}$ lower by 
   4.0--4.5\% on average compared to the NRAO 140 ft at Green Bank and the HI4PI 
   survey.  In several sight lines, GBT measured $N_{\rm HI}$ lower by 10--30\%.
      
\item Including \Htwo\ in the total $N_{\rm H} = N_{\rm HI} + 2N_{\rm H2}$ increases 
    $N_{\rm H}$  by 2--3\% on average at high latitude, with four sight lines
    exhibiting $f_{\rm H2}$ of 7\% to 17\%.  

\item Excluding high velocity gas (HVCs) decreases \Ratio\  by 15\% on average,
and by 40\% for nine sight lines through Complex C.  
    
 \end{itemize}
 
Figure 8 visually illustrates the mean gas-to-dust ratios in our sub-samples. 
For each sub-sample, three points show the shifts that occur when one uses different 
reddening maps (SF11 vs.\ Planck-GN) or sight lines with values of $N_{\rm HI}$ at
all velocities or just low velocity.  Within formal uncertainties, our dust-to-gas ratios 
are consistent with the 8.8 value from Lenz \etal\ (2017) when we exclude the 
low-reddening sight lines, consider gas at all velocities, and use Planck-GN reddening. 
The right-most points in each triplet in Figure 8 (labeled SF11, low) are also consistent
with Lenz \etal\ (2017), but that sub-sample only considers low-velocity gas.  There is
also good evidence that Planck-GN reddening maps are superior to SF11.  The mean 
ratio is sensitive to the AGN sample selection, implying once again that the formal variation 
about the mean underestimates the systematic uncertainties.

As noted, surveys of extragalactic targets at $|b|  > 30^{\circ}$ would be expected to show 
higher gas-to-dust ratios.  However, it is important to assess how much arises from reduced grain 
content, and how that deficiency occurs.  High ratios are clearly seen in sight lines with HVCs 
(18 of 55 of  the AGN sight lines in group~1).  In this sample with both \HI\ and \Htwo\ measurements, 
the mean ratio drops by 15\% when HVCs are excluded.   Some of the anomalously high ratios 
may result from under-estimated reddening when $E(B-V) \leq 0.04$. 
 A comparison of \HI\ column densities obtained from different radio telescopes (Table 2) 
 found differences in $N_{\rm HI}$ between GB and GBT telescopes and from the HI4PI survey, 
 as well as from LAB, LDS, and Effelsberg measurements.  Uncertainties in $E(B-V)$ may be 
 even larger, as converting FIR emission to optical reddening requires precise modeling of 
 dust temperature $T_d$ and grain emissivity index $\beta$.  Appendix C demonstrated the 
 sensitivity of dust radiance to $\beta$ and its anti-correlation with $T_d$.  We also discussed
 the accuracy of $E(B-V)$ in previous FIR studies (SFD98, SF11) compared to values from FIR 
 all-sky maps from the Planck Mission.  Tabulated values of $E(B-V)$ at $|b| > 30\degr$ have 
 systematic uncertainties larger than the variances listed on the IPAC/IRSA website.  Given the 
 sensitivity of the FIR-derived values to grain parameters, the optical extinction may be 
 underestimated in high-latitude AGN sight lines.  
 
 \noindent
We now summarize our survey results for the mean ratios of gas-to-dust,
\begin{enumerate}

\item In the Galactic disk, the mean ratio \Ratio\ of interstellar gas-to-dust in the 
    Galactic disk has been determined as $6.0 \pm 0.2$ (in units of $10^{21}$ 
    cm$^{-2}$~mag$^{-1}$) by many studies (Bohlin \etal\ 1978; Shull \etal\ 2021;  
    Gudennavar \etal\ 2012; Liszt \& Gerin 2023).  For 51 quasars at high Galactic
    latitude, with both \HI\ and \Htwo\ and 
    $0.01 \leq E(B-V) \lesssim 0.1$ (Planck-GN scale), we find mean ratios 
    $10.3\pm0.4$ (gas at all velocities) and $9.2\pm0.3$ (low-velocity). 
       
  \item A portion of the high gas-to-dust ratios likely arises from reduced grain content 
     in HVCs (and some IVCs) owing to low metallicity and shock destruction of grains.  
      For nine sight lines passing through Complex C, with mean $E(B-V) = 0.0151$ (SF11) 
      and 0.0185 (GN), the ratio decreases by 40\% when high velocity gas is excluded.
      
\item Owing to uncertainties in both numerator $N_{\rm HI}$ and denominator $E(B-V)$,
       the gas-to-dust ratio has large and asymmetric errors.  In a comparison of
       36 AGN sight lines, some values of $\log N_{\rm HI}$ differed by 0.05--0.12 (dex) in 
       21-cm observations at various radio telescopes. Compared to data from the GBT 
       100~m ($9.1'$ beam), the average values of $\log N_{\rm HI}$ were higher 
        by $+0.017$ (GB-140~ft, $21'$ beam), $+0.023$ (HI4PI, $16'$ beam), and $+0.059$ 
        Wakker \etal\ (2003).  

\item Elevated $N_{\rm H}/E(B-V)$ may also arise from uncertain FIR estimates of 
    $E(B-V)$, which are sensitive to dust temperature $T_d$ and emissivity index $\beta$.  
    With their observed anti-correlation, $(\beta/\beta_0) = (T_d/T_0)^{-\alpha}$, dust 
    radiance ${\cal R} \equiv \int I_{\nu} \, d \nu$ depends sensitively on the emissivity index, 
    ${\cal R}/ {\cal R}_0 = (\beta/\beta_0)^{-(3 + \beta)/\alpha}$.  For $\alpha \approx 2/3$
    (Martin \etal\ 2012) and $\beta = 1.6 \pm 0.2$ adopted in the 2016 Planck-GN study,
    factor-of-two changes could occur in $E(B-V)$ from variations about the fitted 
    radiance ${\cal R}_0$. 

\item Values of $E(B-V)$ at $|b| > 30\degr$ from Planck-GN dust emission are 
    preferred over those from Planck-DL, SF11, or SFD98.  On average, Planck-GN
    reddening values are 12\% higher than SF11 for $E(B-V) \leq 0.04$, with large
    variations in the (GN/SF11) ratios of $E(B-V)$.  An underestimate of reddening at
    high latitudes and low $E(B-V)$ is consistent with an analysis of Planck data 
    (Casandjian \etal\ 2022), who also found excess dust at high latitude.
     
\end{enumerate}

Reddening maps and variations in the gas-to-dust ratio are important for studies of the ISM 
and many other areas of astrophysics.   Owing to the steep rise of selective extinction toward 
shorter wavelengths, the spectral slopes of de-reddened UV spectra of AGN will be harder 
than found in composite spectra (Stevans \etal\ 2014) which used IPAC/IRSA reddening tables.  
Reddening maps will affect CMB foreground subtraction and derivation of cosmological 
parameters, including ``B-modes" in polarized emission.  Variations in $\log N_{\rm HI}$ 
measurements and FIR-inferred $E(B-V)$ and propagated errors in the numerator and 
denominator result in large, asymmetric errors in \Ratio.  This suggests the need to obtain
high-quality 21-cm observations of a sample of high-latitude quasars to understand the source 
of offsets.  It would also be helpful to update the IPAC/IRSA reddening tables, frequently 
used to convert measured $N_{\rm HI}$ to reddening and extinction.  

\begin{acknowledgements}

 We thank Jean-Marc Cassandjian, Bruce Draine, Charles Danforth, Andrew Fox, 
 Brandon Hensley, and Bart Wakker for helpful discussions, and Harvey Liszt for 
 a thorough and timely referee report.   Jay Lockman and Bart Wakker kindly provided 
 additional 21-cm data from the Green Bank Telescope and expert advice on several
 AGN sight lines.  Marc-Antoine Miville-Desch\^enes provided the optical filter bandpasses 
 used in the Planck analysis.  A portion of this study was supported by the New Horizons 
Mission observational planning for observations of cosmic backgrounds at high 
Galactic latitude.  The authors also acknowledge Interstellar Institute's Program II6 
and the Paris-Saclay University's Institut Pascal for hosting discussions that nourished 
the development of the ideas behind this work.  Our survey made use of the reddening 
website for IPAC/IRSA data of FIR emission and modeled values of dust temperature 
$T_d$ and $E(B-V)$.  

\end{acknowledgements}



\newpage

\appendix

        
\section{Error Propagation in the Ratio $N_{\rm H}$/E(B-V)  }  

Errors in the gas-to-reddening ratio \Ratio\  arise from uncertainties in measurements
of three quantities:  $\log N_{\rm HI}$, $\log N_{\rm H2}$, and $E_{B-V}$, where 
 $N_{\rm H} \equiv N_{\rm HI} + 2 N_{\rm H2}$.  Because $\log N =  (\ln N/2.303)$, we have
 $\sigma_{\log N_H} =  (\sigma_{\ln N_H}  / 2.303) = (\sigma_{N_H} / 2.303 \, N_{\rm H})$. 
The propagated errors on $\log {\rm N}_{\rm H}$  give the weighted formula,
  \begin{equation}
      \sigma^2_{\log N_{\rm H}} = \left( \frac {N_{\rm HI}} {N_{\rm H} } \right)^2 \sigma_{\log N_{\rm HI}}^2 
                                              + \left( \frac {2N_{\rm H2}} {N_{\rm H}} \right)^2 \sigma_{\log N_{\rm H2}}^2 \; .
  \end{equation}  
  Similarly, propagated errors in $N_{\rm H}/E(B-V)$ lead to 
 \begin{equation}
  \Bigl( \frac {\sigma_{\rm ratio}} {{\rm ratio}} \Bigr) ^2  =  \left( \frac { \sigma_{ N_H } } { N_H } \right)^2 + 
                           \left(  \frac { \sigma_{ E_{B-V} } }  { E_{B-V} } \right)^2 
                 =     \Bigl[ 2.303 \,  \sigma_{ \log N_H}  \Bigr] ^2  + 
                      \Bigl[ \frac { \sigma_{E_{B-V} }} {E_{B-V} } \Bigr]^2  \nonumber  \;  .                        
\end{equation} 
We can use this expression to evaluate which errors dominate uncertainty in the gas-to-reddening
ratios discussed in this paper.   Consider a typical example of O-star data (Shull \& Danforth 2019) 
for HD~15137 (O9.5~II-IIIn) 
at $b = -7.58\degr$ with $\log N_{\rm HI} = 21.24\pm0.07$, $\log N_{\rm H2} = 20.27\pm0.05$, 
and $E(B-V) = 0.35 \pm 0.03$.  Here, the molecular fraction $f_{\rm H2} = 0.186$, the total hydrogen 
column density $\log N_{\rm H} = 21.324 \pm 0.058$, and the ratio is
$N_{\rm H}/E(B-V) = (6.02 \pm 0.96) \times10^{21}$ cm$^{-2}$ mag$^{-1}$ with uncertainty
$\sigma_{\rm ratio}/{\rm ratio} = 0.159$ (16\% errors).  Errors on the ratio are dominated by 
uncertainties in $\log N_{\rm HI}$ [0.07 dex or 17\%] and $\log N_{\rm H2}$ (0.05 dex or 
12\%) compared to 9\% error on $E(B-V)$.  Consider now an O-star sight line with lower reddening, 
$E(B-V) = 0.15\pm0.03$ (20\% error) and column densities 
$\log N_{\rm HI} = 21.00\pm0.05$ and $\log N_{\rm H2} = 19.24\pm0.10$.  In this case, 
$\log N_{\rm H} = 21.015 \pm 0.048$ (12\% errors) and the ratio is
$(6.90\pm1.58)\times10^{21}$ with 23\% errors.  In this case, errors on $E(B-V)$ are more important.  

Next, consider a moderately reddened quasar, Mrk~1095 with $\log N_{\rm HI} = 20.969\pm0.07$, 
$\log N_{\rm H2} = 18.76^{+0.21}_{-0.31}$, and $E(B-V) = 0.1099$.  The IPAC/IRSA tables quote an
error of $\pm0.0017$ (1.5\%) on $E(B-V)$ (from SF11), but we suggest a larger systematic error of
$\pm0.02$ (18\%).  From these, we find $\log N_{\rm H} = 20.974\pm0.069$ and 
$\sigma_{\rm ratio}/{\rm ratio} = 0.24$, so that 
$N_{\rm H}/E(B-V) = (6.8 \pm 1.6) \times10^{21}$ cm$^{-2}$ mag$^{-1}$.   Errors on this ratio are 
dominated by uncertainties in both $\log N_{\rm HI}$ (17\%) and $E(B-V)$ (18\%)

As a final example, consider Mrk~279, a high-latitude quasar behind Complex~C with low reddening
(see Table 7) variously quoted as $E(B-V) = 0.0161\pm0.0006$ (SFD98), $0.0138\pm0.0005$
 (SF11), $0.0197\pm0.0063$ (Planck-DL), and $0.0154\pm0.0002$ (Planck-GN).  The column 
 densities are $\log N_{\rm HI} = 20.388\pm0.05$, $\log N_{\rm H2} = 14.42\pm0.09$, and 
 $\log N_{\rm H} = 20.338 \pm 0.048$.  The gas-to-dust ratios (15.8, 11.1, 14.1)are given in the 
 usual ($10^{21}$) units.  The quoted relative errors on $E(B-V)$ are inconsistent with the 26\% 
 dispersion among the three estimates;  systematic errors likely dominate the uncertainty on the ratio.  
 Similar uncertainties (and bias) are likely present in other lightly reddened extragalactic sight lines.
           


\section{Notes on \Htwo\ for Individual AGN Sight Lines}

In Table 1, we listed \Htwo\ column densities for the 55 sight lines in group 1
based on several sources, primarily two 2006 FUSE surveys of high-latitude quasars 
(Gillmon \etal\ 2006; Wakker 2006).  We found generally good agreement, but in 
several cases there were discrepancies. Table 1 includes \Htwo\ column densities for 
38 of the 45 AGN in Gillmon \etal\ (2006).  For 9 other sight lines (H1821+643, 
HE~0226-4110, Mrk~501, Mrk~817, NGC~1399, NGC~3310, NGC~3690, NGC~4214, 
PHL~1811) we adopted values from Wakker (2006), who analyzed additional FUSE 
observations and generated smaller error bars.  Wakker (2006) also provided \Htwo\ 
data for 9 additional targets (H1821+643, HE~0226-4110, Mrk~279, Mrk~501, 
Mrk~817, Mrk~1383, NGC~985, PG~1116+215, Ton~S210).  Table~3 in that paper
also lists line widths (FWHM), which we translate to $b = {\rm FWHM}/1.665$
(assuming a Gaussian profile) to compare with curve-of-growth Doppler parameters
found by Gillmon \etal\ (2006).  Below are notes on several discrepancies and  
our reasons for the selected $N_{\rm H2}$. 

\noindent 
{\bf Fairall~9.}  Richter \etal\ (2001b) detected \Htwo\ with 
$\log N_{\rm H2} = 16.40^{+0.28}_{-0.53}$ in the Magellanic Stream toward Fairall~9 at 
$V_{\rm LSR} = +190$~\kms.   No \Htwo\ was seen at low velocity (Gillmon \etal\ 2006).  
We quote the high-velocity value, $\log N_{\rm H2} = 16.40$ in Table~1.

\noindent
{\bf H1821+643.}  Published \Htwo\ column densities and Doppler parameters 
differ between two FUSE surveys:
$\log N_{\rm H2} = 17.91^{+0.13}_{-0.20}$ with $b = 1.7^{+0.8}_{-0.7}$~\kms\ 
(Gillmon \etal\ 2006) and $\log N_{\rm H2} = 15.99^{+0.16}_{-0.06}$ based on 
$b = 4.3$~\kms\  inferred from FWHM = 7.2~\kms\ quoted in Table 3 of Wakker (2006).  
The difference arises from the two curve-of-growth $b$-values (or FWHM) fitted 
to the data.  We adopted the lower columns of Wakker (2006) because his higher 
Doppler parameter seemed more plausible physically.      

\noindent
{\bf HE~0226-4110.}  Gillmon \etal\ (2006) quoted $\log N_{\rm H2} < 14.29$, but 
Fox \etal\ (2005) detected $\log N_{\rm H2} = 14.58\pm0.09$.  Two subsequent 
analyses quoted 14.54 (Lehner \etal\ 2006) and $14.56^{+0.15}_{-0.10}$ (Wakker 2006).  
Summing the column densities $N(J)$ for $J =$ 0--3 in Table 3 of Wakker (2006), we 
find $\log N_{\rm H2} = 14.60 \pm 0.14$, with propagation of errors on $N(J)$.  
We use the latter value in Table~1.

\noindent
{\bf MRC 2251-178.}  Gillmon \etal\ (2006) quoted
$\log N_{\rm H2} = 14.54^{+0.23}_{-0.17}$, based on a detection in $J = 1$ 
and an upper limit in $J = 0$.  Wakker (2006) listed a total column density
$\log N_{\rm H2} = 15.02^{+0.45}_{-0.23}$, but that included upper limits for 
column densities in $J =$ 2--4, which should probably not be included in the 
total sum. Instead, we adopt $\log N_{\rm H2} = 14.87^{+0.23}_{-0.17}$ from 
the detected column densities in $J = 0$ and $J = 1$.

\noindent
{\bf Mrk 106.}  Gillmon \etal\ (2006) quoted $\log N_{\rm H2} = 16.23^{+0.21}_{-0.15}$
with $b = 14.9^{+3.5}_{-2.6}$~\kms.  Wakker (2006) quoted 
$\log N_{\rm H2} = 18.54^{+0.11}_{-0.10}$ with FWHM = 7.5~\kms\ ($b = 4.5$~\kms).  
We adopt the smaller column density, since the high value in Wakker (2006) should
produce detectable damping wings.  (However, the FUSE data do not have high S/N.)
Another argument for the lower value is that a large \Htwo\ column density and high
molecular fraction ($f_{\rm H2} \approx 0.03$) would be unlikely at the estimated 
$E(B-V) \approx 0.0235$ (SF11).  

\noindent
{\bf Mrk 279.}  This target did not appear in Gillmon \etal\ (2006).  Wakker (2006)
mentioned no low-velocity \Htwo\ absorption, but his Table 4 listed 
$\log N_{\rm H2} = 14.42\pm0.09$ in the LLIV arch at $V_{\rm LSR} = -41$~\kms.  
We quote the latter (IVC) value in Table 1.  

\noindent
{\bf Mrk 501.}  We adopted the value $\log N_{\rm H2} = 15.49^{+0.66}_{-0.24}$ from
Wakker (2006), as it was based on additional data than used by Gillmon \etal\ (2006).  

\noindent
{\bf Mrk 817.}  We adopted the value $\log N_{\rm H2} = 14.48^{+0.08}_{-0.07}$ from
Wakker (2006), as it was based on additional data than used by Gillmon \etal\ (2006).  

\noindent
{\bf Mrk 876.}  Gillmon \etal\ (2006) quoted $\log N_{\rm H2} = 16.58$ (low-velocity) 
and 15.75 (IVC at $-33$~\kms) both seen in $J =$ 0--3.  The fitted Doppler parameters 
were $b = 7.3$~\kms\ and 5.7~\kms\ respectively.  Wakker (2006) found a much higher 
column density for the low-velocity component,
$\log N_{\rm H2} = 18.07^{+0.22}_{-0.22}$ with FWHM = 7.6~\kms\ ($b = 4.6$~\kms),
but a similar value to Gillmon \etal\ (2006), $15.47^{+0.39}_{-0.13}$, for the IVC component.  
We adopt the Gillmon \etal\ (2006) values, since the high value in Wakker (2006) 
might produce detectable damping wings. (The FUSE data do not have high S/N.)
Such a large molecular fraction ($f_{\rm H2} \approx 0.014$) would be unlikely at 
the estimated $E(B-V) \approx 0.0230$ (SF11).  

\noindent
{\bf Mrk 1383.}  We adopted the value $\log N_{\rm H2} = 14.78^{+0.20}_{-0.14}$ from
Wakker (2006), as it was based on additional data than used by Gillmon \etal\ (2006)
and had smaller errors on $N(J)$.  

\noindent
{\bf NGC 985.}  The two FUSE surveys find similar values, 
$\log N_{\rm H2} = 16.05^{+1.95}_{-0.33}$ (Gillmon \etal\ 2006) and
$\log N_{\rm H2} = 16.07^{+0.77}_{-0.24}$ (Wakker 2006).  We list the latter value 
(smaller errors).  

\noindent
{\bf PG 1116+215.}  Gillmon \etal\ (2006) quoted upper limits on low-velocity \Htwo\ 
in $J = 0$ and $J = 1$, with $\log N(0) < 13.78$ and $\log N(1) < 13.93$.  Wakker (2006) 
quoted an iupper limit $\log N(0) < 13.79$ but detected 
$\log N(1) = 13.82^{+0.20}_{-0.22}$ in the low-velocity component.  He also quoted \Htwo\ 
detections in the S1 IVC  (-44~\kms) with $\log N_{\rm H2} = 15.27^{+0.44}_{-0.23}$.  
However, summing the column densities $N(J)$ for $J =$ 0--3 in Wakker (2006), 
we find $\log N_{\rm H2} = 16.01^{+0.38}_{-0.27}$, with errors propagated $\log N(J)$.  
We quote the latter (IVC) value in Table 1.

\noindent
{\bf Ton~S210.}  The FUSE surveys find different values, 
$\log N_{\rm H2} = 16.57^{+1.18}_{-1.38}$ (Gillmon \etal\ 2006) with 
$b = 4.4^{+5.6}_{-2.4}$~\kms\ and 
$\log N_{\rm H2} = 15.61^{+0.32}_{-0.20}$ (Wakker 2006) with FWHM = 7.3~\kms\
($b = 4.4$~\kms).  We list the latter in Table 1 (additional data analyzed).



\section{Sensitivity of $E(B-V)$ Estimates to Dust Modeling}  

The FIR-inferred values of optical selective extinction $E(B-V)$ are derived from a chain of 
computations and correlations with dust optical depth or integrated emission intensity.
Figure 14 of Lenz \etal\ (2017) is instructive, showing systematic differences between reddening 
maps as a function of dust temperature.  The values from Planck Collaboration papers
 (2011, 2014, 2016, 2020) employ parameterized models of thermal dust emission based on 
 a modified blackbody (MBB) model with emissivity $\epsilon(\nu) \propto \nu^{\beta}$.  
The parameters are based on assumed grain composition (e.g., graphite + silicates), size 
distributions, and emission properties (e.g., Draine \& Li 2007; Compi\`egne \etal\ 2011).
Most past studies assumed a single-temperature dust distribution, with $T_d \approx$ 16--23~K 
and $\beta \approx$ 1.5--2.0.   Meisner \& Finkbeiner (2015) explored a two-component model 
with ``hot dust" ($T_d \approx 16.2$~K, $\beta = 1.67$) and ``cold dust" 
($T_d \approx 9.15$~K, $\beta = 2.70$).  

 The observed determinations of the $\beta$-index (Table 4) illustrate the difficulties in dust 
 modeling.  An early study (Planck Collaboration XXIV 2011) used $\beta = 1.8$, noting 
 compatibility with the FIRAS spectrum of the diffuse ISM. Subsequent Planck papers found
 $\beta \approx 1.6$ for high-latitude sight lines.  Their mean and $1\sigma$ variance were 
 $\beta = 1.59 \pm 0.12$ (Planck Collaboration XI 2014) and $\beta = 1.63 \pm 0.17$ 
 (Planck Collaboration XLVIII  2016).  Several of these studies used a ``dust radiance" method  
 to estimate the thermal dust emission in a MBB model, with radiance defined as
\begin{equation}
   { \cal R} = \int I_{\nu} \, d \nu = \int \tau_{353} \, B_{\nu}(T_d)
         \left( \frac {\nu} {353~{\rm GHz} } \right)^{\beta} \, d \nu 
      \propto \tau_{353} \, T_d^{4+\beta}  \; .
 \end{equation}
Here $\tau_{353}$ is the dust optical depth at  $\nu_0 = 353$~GHz, the frequency at which 
 the emissivity index is normalized.   In the MBB formulation, with $\tau_{\nu} = I_{\nu}/B_{\nu}$
 and $B_{\nu}(T_d) = (2 \nu^2 T_d / c^2)$ in the Rayleigh-Jeans limit, the dust optical depth 
 $\tau_{353} \propto T_d^{-1}$.  Because of the sensitivity of dust emissivity to temperature, 
 it is useful to understand how the inferred radiance depends on the assumed index $\beta$.  
 Several papers (see Figure 16 in Planck Collaboration XI 2014) found an anti-correlation
 between $\beta_{\rm obs}$ and $T_{\rm obs}$.  A possible explanation is that when dust 
 emits more efficiently it acquires a lower temperature.  
 
 Martin \etal\ (2012) proposed an approximate relation, 
 $(\beta/1.8) \approx (T_d/17.9~{\rm K})^{-2/3}$, based on 100--500~$\mu$m Galactic-plane 
 data from Paradis \etal\ (2010).  The latter paper fitted to a general relation,
 $\beta \propto T_d^{-\alpha}$ and $\alpha \approx 4/3$ (with substantial scatter).  We adopt 
 the formulation $(\beta/\beta_0) = (T_d/T_0)^{-\alpha}$, where $T_0$ and $\beta_0$ are fiducial 
 parameters chosen at the center of the distributions.  Combining the scaling of 
 ${\cal R} \propto \tau_{353} T_d^{4+\beta} \propto T_d^{3 +\beta}$ with the anti-correlation 
 ($T_d \propto \beta^{-1/\alpha}$) between dust emission index and dust temperature, we find 
 that radiance is quite sensitive to changes in these indices ($\alpha$ and $\beta$),
 \begin{equation} 
     \frac {{\cal R}} {{\cal R}_0} = \left( \frac {\beta} {\beta_0} \right)^{-(3 + \beta)/\alpha}   \;  .
\end{equation} 
For the range of emission indices, $\beta = 1.6 \pm 0.2$, adopted in the 2016 Planck-GN
study (and $\alpha = 2/3$) the radiance factor changes by a factor of 2.3 about ${\cal R}_0$.
Over the range of adopted indices (Table 5) the differences could be even larger.



\begin{deluxetable} {lrrl cclc ccc}
\tablecolumns{11}
\tabletypesize{\scriptsize}
\tablenum{1}
\tablewidth{0pt}
\tablecaption{Data on 94 AGN Sight Lines and their ``Gas-to-Dust" Ratios\tablenotemark{a} }  

\tablehead{
\colhead{ AGN Name } 
& \colhead{ $\ell$ }
& \colhead{ $b$ } 
& \colhead{$E(B-V)$ }
& \colhead{ $\log N_{\rm HI}$\tablenotemark{b}  } 
& \colhead{ $\log N_{\rm HI}$\tablenotemark{b}  } 
& \colhead{ $\log N_{\rm H2}$\tablenotemark{b} } 
& \colhead{ $\log N_{\rm H}$\tablenotemark{b} }
& \colhead{ Ratio\tablenotemark{c} }  
& \colhead{ Ratio\tablenotemark{c} }  
& \colhead{ GN-Ratio\tablenotemark{c} }
\\
\colhead{ } 
& \colhead{ (deg) } 
& \colhead{ (deg) }
& \colhead{(SF11)}
& \colhead{ (all-v)}  
& \colhead{ (low-v)}  
& \colhead{(FUSE)} 
& \colhead{ (total) } 
& \colhead{(all-v) } 
& \colhead{(low-v) }
& \colhead{(all-v) } 
 \\
\colhead{(1)}
& \colhead{(2)}  
& \colhead{(3)}  
& \colhead{(4)}  
& \colhead{(5)}  
& \colhead{(6)}  
& \colhead{(7)}  
& \colhead{(8)}  
& \colhead{(9)}
& \colhead{(10)}  
& \colhead{(11)}
}

\startdata

3C~249.1          & 130.39 &  38.55 & $0.0301\pm0.0020$ & 20.446 & 20.335      & $18.98^{+0.12}_{-0.14}$   & 20.475       & 9.92 & 7.82 & 7.48  \\
3C~273             & 289.95 &  64.36 & $0.0179\pm0.0004$ & 20.222 & 20.222      & $15.72^{+0.13}_{-0.09}$   & 20.222       & 9.32 & 9.32 & 8.02  \\     
ESO~141-G55  & 338.18 & -26.71 & $0.0944\pm0.0062$ & 20.788 & 20.704      & $19.32^{+0.07}_{-0.07}$   & 20.817       & 6.95 & 5.80 & 7.05  \\   
Fairall~9            & 295.07 & -57.83 & $0.0217\pm0.0011$ & 20.526 & 20.383      & $16.40^{+0.28}_{-0.53}$   & 20.526       & 15.5 & 11.1 & 15.5  \\
H1821+643       &  94.00  &  27.42 & $0.0370\pm0.0007$ & 20.584 & 20.570      & $15.99^{+0.16}_{-0.06}$   & 20.584       & 10.4 & 10.0 & 10.0  \\    
HE~0226-4110  & 253.94 & -65.77 & $0.0132\pm0.0005$ & 20.272 & 20.272      & $14.60^{+0.14}_{-0.14}$   & 20.272       & 14.2 & 14.2 & 11.6 \\                    
HE~1143-1810  & 281.85 &  41.71 & $0.0331\pm0.0005$ & 20.506 & 20.496      & $16.54^{+1.32}_{-0.68}$   & 20.506       & 9.69 & 9.46 & 9.12 \\  
HS~0624+6907 &145.71 &  23.35  & $0.0845\pm0.0021$ & 20.898 & 20.798      & $19.82^{+0.09}_{-0.09}$   & 20.964       & 10.9 & 8.97 & 8.66 \\   
MRC~2251-178 &  46.20 & -61.33 &$0.0335\pm0.0011$  & 20.415 & 20.407       & $14.54^{+0.23}_{-0.17}$   & 20.415       & 7.77 & 7.61 & 10.3 \\   
Mrk~9                & 158.36 &  28.75 &$0.0503\pm0.0015$ & 20.677 & 20.639        & $19.36^{+0.09}_{-0.08}$   & 20.717       & 10.4 & 9.57 & 7.88 \\   
Mrk 106             & 161.14 &  42.88 &$0.0235\pm0.0010$ & 20.453 & 20.349        & $16.22^{+0.21}_{-0.15}$   & 20.453       & 12.1 & 9.50 & 9.13 \\                              
Mrk 116             & 160.53 &  44.84 & $0.0292\pm0.0028$ & 20.505 & 20.429       & $19.09^{+0.08}_{-0.09}$   & 20.537       & 11.8 & 10.0 & 8.30 \\                
Mrk 205             & 125.45 &  41.67 & $0.0344\pm0.0011$ & 20.508 & 20.405       & $16.53^{+0.13}_{-0.37}$   & 20.508       & 9.37 & 7.38 & 8.54 \\    
Mrk 209             & 134.15 &  68.08 & $0.0122\pm0.0005$ & 20.052 & 20.006       & $<14.48$                          & 20.052       & 9.24 & 8.30$^*$ & 8.35 \\          
Mrk 279             & 115.04 & 46.86  & $0.0138\pm0.0005$ & 20.338 & 19.926       & $14.42^{+0.09}_{-0.09}$   & 20.338       & 15.8 & 6.11 & 14.2  \\                                                            
Mrk 290             &   91.49 &  47.95 & $0.0120\pm0.0008$ & 20.426 & 20.125       & $16.18^{+0.49}_{-0.39}$   & 20.426      & 22.2 & 11.1 & 16.4 \\          
Mrk 335             & 108.76 & -41.42 & $0.0305\pm0.0031$ & 20.567 & 20.428       & $18.83^{+0.08}_{-0.08}$   & 20.583      & 12.5 & 9.23 & 9.56 \\                          
Mrk 421             & 179.83 &  65.03 & $0.0130\pm0.0011$ & 20.166 & 19.936       & $14.63^{+0.09}_{-0.10}$    & 20.166      & 11.3 & 6.64 & 11.5  \\          
Mrk 478             &   59.24 &  65.03 & $0.0111\pm0.0009$ & 20.010 & 19.963       & $<14.56$                            & 20.010      & 9.22 & 8.28 & 9.74  \\     
Mrk 501             &   63.60 &  38.86 & $0.0164\pm0.0003$ & 20.314 & 20.242      & $15.49^{+0.66}_{-0.24}$     & 20.314      & 12.6 & 10.6 & 11.3 \\     
Mrk 509             &   35.97 & -29.86 & $0.0492\pm0.0005$ & 20.613 & 20.577      & $17.87^{+0.31}_{-0.78}$    & 20.614       & 8.36 & 7.71 & 8.36 \\     
Mrk 817             & 100.30 &  53.48 & ${\bf 0.0059\pm0.0002}$ & 20.085 & 19.950 & $14.48^{+0.08}_{-0.07}$  & 20.085       & 20.6 & 15.1$^*$ & 10.6 \\                        
Mrk 876             &   98.27 &  40.38 & $0.0230\pm0.0009$ & 20.424 & 20.210      & $16.58^{+1.96}_{-0.42}$    & 20.424      & 11.5 & 7.05 & 9.14 \\     
Mrk 1095           & 201.69 & -21.13 & $0.1099\pm0.0017$ & 20.969 & 20.969      & $18.76^{+0.21}_{-0.31}$    & 20.975      & 8.62 & 8.62 & 9.96 \\     
Mrk 1383           & 349.22 &  55.12 & $0.0276\pm0.0009$ & 20.547 & 20.547      & $14.78^{+0.20}_{-0.14}$    & 20.547      & 12.8 & 12.8 & 15.2 \\     
Mrk 1513           &   63.67 & -29.07 & $0.0370\pm0.0017$ & 20.555 & 20.525      & $16.42^{+1.08}_{-0.26}$    & 20.555       & 9.70 & 9.06 &7.98 \\    
MS~0700.7+6338 & 152.47 &25.63& $0.0448\pm0.0014$ & 20.609 & 20.609     & $18.75^{+0.27}_{-0.68}$    & 20.620        & 9.31 & 9.31 & 8.48 \\                        
NGC~985          & 180.84 & -59.49 & $0.0282\pm0.0003$ & 20.522 & 20.522     & $16.07^{+0.77}_{-0.33}$     & 20.522        & 11.8 & 11.8 & 13.6 \\                                       
NGC~1068        & 172.10 & -51.93 & $0.0289\pm0.0004$ & 20.436 & 20.416     & $18.13^{+0.13}_{-0.17}$     & 20.440        & 9.53 & 9.11 & 8.56 \\       
NGC~1399        & 236.72 & -53.63 & $0.0109\pm0.0004$ & 20.196 & 20.196     & $<14.55$                            & 20.196        & 14.4 &14.4 & 10.7 \\                                          
NGC~1705        & 261.08 & -38.74 & ${\bf 0.0070\pm0.0005}$ & 20.251 & 20.119 & $<14.17$                         & 20.251        & 25.5 & 18.8 & 18.7  \\    
NGC~3310        & 156.60 &  54.06 & $0.0192\pm0.0005$ & 20.150 & 19.888     & $19.14^{+0.22}_{-0.42}$    & 20.227         & 8.78 & 8.78$^*$ & 6.08 \\            
NGC~3690        & 141.91 &  55.41 & $0.0144\pm0.0002$ & 19.994 & 19.758     & $<14.40$                            & 19.994         & 6.85 & 6.47$^*$ & 6.37 \\        
NGC~4151        & 155.08 &  75.06 & $0.0237\pm0.0011$ & 20.322 & 20.284     & $16.70^{+0.93}_{-0.31}$     & 20.322         & 8.86 & 8.86$^*$ & 11.3 \\    
NGC~4214        & 160.24 &  78.07 & $0.0187\pm0.0003$ & 20.232 & 19.538     & $15.22^{+0.31}_{-0.19}$     & 20.232         & 9.12 & 9.12$^*$ & 11.2 \\   
NGC~4670        & 212.69 &  88.63 & $0.0128\pm0.0003$ & 20.040 & 19.951     & $14.72^{+0.13}_{-0.16}$     & 20.040         & 8.56 & 6.98 & 9.79 \\                      
NGC~7469        &   83.10 & -45.47 & $0.0599\pm0.0015$ & 20.649 & 20.646     & $19.67^{+0.10}_{-0.10}$     & 20.732         & 9.00 & 8.94 & 7.59 \\     
NGC~7714        &   88.22 & -55.56 & $0.0451\pm0.0002$ & 20.674 & 20.636     & $18.94^{+0.05}_{-0.05}$     & 20.690         & 10.9 & 10.0 & 8.20 \\       
PG~0804+761   & 138.28 &  31.03 & $0.0302\pm0.0005$ & 20.588 & 20.545     & $18.66^{+0.14}_{-0.19}$     & 20.598         & 13.1 & 11.9 & 10.4  \\              
PG~0844+349   & 188.56 &  39.97 & $0.0314\pm0.0007$ & 20.475 & 20.475     & $18.22^{+0.18}_{-0.28}$     & 20.480         & 9.62 & 9.62 & 11.3 \\     
PG~0953+414   & 179.79 &  51.71 & $0.0102\pm0.0006$ & 20.087 & 20.004     & $15.03^{+0.11}_{-0.10}$     & 20.087         & 12.0 & 9.90 & 10.1 \\                        
PG~1116+215   & 223.36 &  68.21 & $0.0193\pm0.0001$ & 20.075 & 19.705      & $16.01^{+0.44}_{-0.33}$    & 20.075         & 6.16 & 6.16$^*$ & 14.1 \\          
PG~1211+143   & 267.55 &  74.32 & $0.0286\pm0.0011$ & 20.421 & 20.249      & $18.38^{+0.15}_{-0.14}$    & 20.429         & 9.39 & 9.39$^*$ & 12.6  \\     
PG~1259+593   & 120.56 &  58.05 & ${\bf 0.0070\pm0.0004}$ & 20.225 &19.669 & $14.75^{+0.10}_{-0.12}$  & 20.225         & 24.0 & 11.2$^*$ &19.5  \\               
PG~1302-102   & 308.59 &  52.16 & $0.0376\pm0.0009$ & 20.502 & 20.368       & $15.62^{+1.41}_{-0.16}$   & 20.502         & 8.44 & 8.41 & 8.69 \\            
PG~1351+640  &  111.89 &  52.02 & $0.0177\pm0.0003$ & 20.441 & 20.286       & $18.34^{+0.20}_{-0.11}$   & 20.448         & 15.9 & 11.2$^*$ & 18.0  \\       
PG~1626+554   &  84.51 & 42.19  & ${\bf 0.0053\pm0.0005}$ & 20.053 &19.936 & $15.14^{+0.54}_{-0.20}$   & 20.053         & 21.3 & 16.3 & 8.24  \\       
PHL~1811         &   47.46 & -44.81 & $0.0415\pm0.0031$ & 20.592 & 20.592       & $19.36^{+0.07}_{-0.06}$  & 20.640          & 10.5 &10.5 & 6.10   \\               
PKS~0405-12   & 204.93 & -41.76 & $0.0503\pm0.0036$ & 20.537 & 20.537        & $15.79^{+0.25}_{-0.12}$  & 20.537         & 6.85 & 6.85 & 8.54 \\       
PKS~0558-504 & 257.96 & -28.57 & $0.0388\pm0.0012$ & 20.675 & 20.526        & $15.44^{+0.18}_{-0.12}$  & 20.675         & 12.2 & 8.66 & 9.92  \\         
PKS~2005-489 & 350.37 & -32.60 & $0.0479\pm0.0007$ & 20.655 & 20.600        & $15.07^{+0.10}_{-0.10}$  & 20.655         & 9.43 & 8.31& 12.9  \\      
PKS~2155-304 &   17.73 & -52.25 & $0.0185\pm0.0005$ & 20.123 & 20.123        & $14.17^{+0.11}_{-0.14}$  & 20.123          & 7.17 & 7.17 & 12.5 \\      
Ton~S180         & 139.00 & -85.07 & $0.0123\pm0.0002$ & 20.097 & 20.083        & $<14.37$                          & 20.097          & 10.2 & 9.84 & 13.8  \\      
Ton~S210         & 224.97 & -83.16 & $0.0144\pm0.0004$ & 20.205 & 20.193        & $15.61^{+0.32}_{-0.20}$  & 20.205          & 11.1 & 10.8 & 12.8  \\      
VII~Zw~118      & 151.36 &  25.99 & $0.0330\pm0.0008$ & 20.558 & 20.558        & $18.84^{+0.10}_{-0.12}$  & 20.574          & 11.4 & 11.4 & 8.58  \\      
                         &             &            &                                  &             &                    &                                          &                      &         &         &      \\ 
ESO350-IG38  & 328.06 & -82.85  & ${\bf 0.0096\pm0.0001}$ & 20.201 & 20.157 &    NA                                & 20.201          & 16.6 & 15.0  & 17.0 \\      
ESO572-G34   & 286.12 &  42.12  & $0.0340\pm0.0004$ & 20.410 & 20.410         &    NA                               & 20.410           & 7.56  & 7.56 & 6.54 \\      
HE~0238-1904 & 200.48 & -63.63 & $0.0272\pm0.0011$ & 20.417 & 20.368         &   NA                                & 20.417           & 9.61  & 8.58  & 8.97  \\
Mrk 487            &   87.84 &  49.03 & $0.0120\pm0.0003$ & 20.142 & 20.094         &   NA                                & 20.142           & 11.6  & 10.3  & 8.66  \\
NGC~4649       & 295.88 &  74.34 & $0.0226\pm0.0006$ & 20.297 & 20.297         &   NA                                & 20.297           & 8.77  & 8.77  & 12.2  \\
HE~1228+0131& 291.26 & 63.66 & $0.0162\pm0.0014$ & 20.230 & 20.181          &   NA                                & 20.230           & 10.5  & 9.37  & 10.1  \\
HE~1326-0516 & 320.07 & 56.07 & $0.0256\pm0.0006$ & 20.331 & 20.308          &   NA                                & 20.331           & 8.37  & 7.94  & 7.88  \\
HS~1102+3441 & 188.56 & 66.22 & $0.0197\pm0.0012$ & 20.189 & 19.814         &   NA                                & 20.189           & 7.85  & 7.85$^*$ & 9.33  \\
Mrk~36             & 201.36 & 66.49  & $0.0250\pm0.0012$ & 20.302 & 20.123         &   NA                                & 20.302           & 8.01  & 5.31 & 8.79  \\
Mrk~54             & 110.64 & 84.55  & $0.0129\pm0.0004$ & 20.110 & 20.110          &   NA                                & 20.110           & 10.0  & 10.0 & 8.96  \\
Mrk~59             & 111.54 & 82.12  & ${\bf 0.0093\pm0.0004}$ & 20.016 &19.920   &   NA                                & 20.016           & 11.2 & 11.2$^*$ & 8.16  \\
Mrk~487           &  82.84 & 49.03  & $0.0120\pm0.0003$ & 20.142 & 20.094          &   NA                                & 20.094           & 11.6  & 10.3 & 9.33  \\
Mrk~734           & 244.75 & 63.94 & $0.0267\pm0.0011$ & 20.385 & 20.251          &   NA                                & 20.385           & 9.09  & 9.09$^*$ & 9.79   \\
Mrk~771           & 269.44 & 81.74 & $0.0233\pm0.0003$ & 20.377 & 20.159          &   NA                                & 20.377           & 10.2  & 6.19 & 6.12  \\
Mrk~829           &   58.76 & 63.25 & $0.0108\pm0.0003$ & 19.977 & 19.977          &   NA                                & 19.977           & 8.79  & 8.79  & 7.99  \\
Mrk~926           &  64.09 & -58.76 & $0.0354\pm0.0007$ & 20.436 & 20.366          &   NA                                & 20.436           & 7.70  & 6.56 & 10.9   \\
Mrk~1502         & 123.75 & -50.18 & $0.0559\pm0.0018$ & 20.669 & 20.669         &   NA                                & 20.669           & 8.35  & 8.35 & 6.76   \\
NGC~3504       & 204.60 & -66.04 & $0.0228\pm0.0002$ & 20.294 & 19.628         &   NA                                & 20.294           & 8.63  & 8.63$^*$ & 10.2  \\
NGC~3991       & 185.68 &  77.20 & $0.0189\pm0.0002$ & 20.217 &19.966          &   NA                                & 20.217           & 8.72  & 8.72$^*$ & 9.73 \\ 
NGC~5548       &   31.96 & 70.50 & $0.0168\pm0.0009$ & 20.203 &20.203           &   NA                                & 20.203           & 9.49  & 9.49 & 9.16   \\ 
NGC~7496       & 347.84 & -63.80 & ${\bf 0.0084\pm0.0001}$ & 20.133 & 20.114  &   NA                               & 20.133           & 16.2  & 15.5  & 13.6  \\ 
PG~0947+396  & 182.85 & 50.75 & $0.0162\pm0.0006$ & 20.224 & 20.079          &  NA                                & 20.224           & 10.3  & 7.40  & 9.01 \\   
PG~1001+291  &200.09 & 53.20  & $0.0190\pm0.0005$ & 20.249 & 19.801          &  NA                                & 20.249           & 9.33  & 9.33$^*$ & 8.83  \\
PG~1004+130  &225.12 & 49.12  & $0.0331\pm0.0005$ & 20.569 & 20.569          &  NA                                & 20.569           & 11.2  & 11.2 & 9.21  \\
PG~1048+342  &190.60 & 63.44  & $0.0199\pm0.0012$ & 20.202 & 19.958          &  NA                                & 20.202           & 8.00  & 8.00$^*$  & 7.86  \\
PG~1216+069  & 281.07 & 68.14 & $0.0186\pm0.0007$ & 20.209 & 20.209          &  NA                                & 20.209           & 8.69  & 8.69 & 8.01  \\
PG~1307+085  & 316.79 & 70.71 & $0.0292\pm0.0008$ & 20.341 & 20.324          &  NA                                & 20.341           & 7.51  & 7.23 & 7.87   \\
PG~1352+183  &   4.37 & 72.87  & $0.0159\pm0.0006$ & 20.223 & 20.223           &  NA                                & 20.223           & 10.5  & 10.5  & 10.7 \\
PG~1402+261  & 32.96 & 73.46  & $0.0132\pm0.0004$ & 20.158 & 20.158           &  NA                                & 20.158           & 10.9  & 10.9  & 9.89  \\
PG~1404+226  & 21.48 & 72.37  & $0.0190\pm0.0003$ & 20.309 & 20.271           &  NA                                & 20.309           & 10.7  & 9.82  & 9.96  \\
PG~1411+442  & 83.83 & 66.35  & ${\bf 0.0084\pm0.0012}$ & 19.880 & 19.857    &  NA                                & 19.880           & 9.04 & 8.57 & 6.54  \\
PG~1415+451  &  84.72 & 65.32 & ${\bf 0.0071\pm0.0003}$ & 19.816 & 19.739    &  NA                                & 19.816          & 9.23  & 7.72 & 6.14  \\ 
PG~1444+407   & 69.90 & 62.72 & $0.0114\pm0.0004$ & 20.031 & 19.866           &  NA                                 & 20.031          & 9.43  & 6.45 & 6.69  \\
PG~2349-014    & 91.66 & -60.36 & $0.0234\pm0.0004$ & 20.483 & 20.478          &  NA                                & 20.483          & 13.0  & 12.9 & 9.37  \\
SBS~0335-052  & 191.34 & -44.69 & $0.0402\pm0.0015$  & 20.556 & 20.556       &  NA                                & 20.556          & 8.94  & 8.94 & 7.40  \\
SBS~1415+437 & 81.96 & 66.20 & ${\bf 0.0077\pm0.0003}$ & 20.021 & 19.920    &  NA                                & 20.021          & 13.6 & 10.8  & 9.39 \\
Tol~0440-381    & 241.07 & -40.98 & $0.0130\pm0.0005$ & 20.331 & 20.331         &  NA                                & 20.331          & 16.5 & 16.5 & 13.0  \\
Ton~1187          & 188.33 & 55.38 & ${\bf 0.0093\pm0.0008}$ & 20.076 & 19.862   &  NA                                & 20.076          & 12.8 &  7.83 & 12.0 \\
vZ~1128            &  42.50 & 78.68 & $0.0117\pm0.0005$  & 20.037 & 20.037          &  NA                                & 20.037           & 9.30  & 9.30 & 9.36  \\

\enddata

\tablenotetext{a} {AGN sight lines surveyed in H~I  21-cm emission (Wakker \etal\ 2003).    
    Values of $\log N_{\rm HI}$ for NGC~1068 are correct; they were incorrectly transcribed
    in Gillmon \etal\ (2006).  The first 55 AGN also have \Htwo\ column densities measured with 
    FUSE ultraviolet spectra, with some survey duplications: 
    38 in Gillmon \etal\ (2006), 18 in Wakker (2006),  8 in Collins \etal\ (2003), one (Fairall~9) 
    in Richter \etal\  (2001b) and one (HE~0226-4110) in both Fox \etal\ (2005) and Wakker \etal\ (2006).
    See Appendix~B for more details.  The next 39 AGN only have H~I measurements.  Columns 1--3 
    provide AGN name and Galactic coordinates.   Column~4 gives the FIR-inferred $E(B-V)$ from 
    SF11, as tabulated on the IRAC/IRSA website. Column~5 lists H~I column densities ($\log N_{\rm HI}$) 
    for all velocities (Wakker \etal\ 2003), and column 6 gives H~I excluding HVCs and most IVCs. Column~7 
    gives \Htwo\ column densities, and column 8 gives values of $N_H = N_{\rm HI} + 2 N_{\rm H2}$ at all 
    velocities.  The last three columns present ratios $N_H/ E(B-V)$ in units 
    $10^{21}~{\rm cm}^{-2}~{\rm mag}^{-1}$.   
    Column 9 presents ratios for gas at all velocities and SF11 reddening.  Column 10 shows ratios for
    low-velocity gas (SF11 reddening) except when marked with an asterisk for inclusion of strong IVCs.  
    Column 11 gives the all-velocities ratio, using $E(B-V)$ from Planck-GN. Ratios for 11 
    sight lines with $E(B-V)$ shown in boldface have uncertain $E(B-V) < 0.01$ (SF11). }
    
 \tablenotetext{b} {Column 5 lists low-velocity values after subtracting HVCs and most IVCs.  Several sight
   lines pass through strong IVCs (IV~Arch, IV18, IV19, IV26, S1) some of which are important contributors 
   to the total H~I (see notes in text).  With \Htwo\ detected in absorption they likely contain dust.  
  The gas-to-dust ratios for these sight lines are computed with N$_{\rm HI}$ that includes the IVCs, marked
  with asterisks (Column 10).  Most high-latitude sight lines exhibit ratios higher than the mean value, 
  $\langle N_H/E(B-V) \rangle \approx 6 \times 10^{21}$ cm$^{-2}$ mag$^{-1}$, measured toward OB stars 
  in the Galactic disk (Bohlin \etal\ 1978; Shull \etal\ 2021). }
 
 \tablenotetext{c} {We list three values of $N_H/ E(B-V)$, two with $E(B-V)$ from SF11 (columns 9 and 10) and
    one using Planck~GN (column~11).   Columns 9 and 11 include gas at all velocities, while column~10 uses
    gas at low velocities only. In 17 cases, noted by asterisks, we included strong IVCs with the low velocity gas.  }
  
\end{deluxetable}



\begin{deluxetable} {lccc ccccc}
\tablecolumns{9}
 \tabletypesize{\scriptsize}

\tablenum{2}
\tablewidth{0pt}
\tablecaption{Comparison of H~I Column Densities\tablenotemark{a} }  

\tablehead{
\colhead{ AGN Name } 
& \colhead{ $\log N_{\rm HI}$ } 
& \colhead{ $\log N_{\rm HI}$ } 
& \colhead{ $\log N_{\rm HI}$ } 
& \colhead{ $\log N_{\rm HI}$ } 
& \colhead{ $\log N_{\rm HI}$ } 
& \colhead{ $\Delta \log N_{\rm HI}$ } 
& \colhead{ $\Delta \log N_{\rm HI}$ } 
& \colhead{ $\Delta \log N_{\rm HI}$ } 
\\
   \colhead{ } 
& \colhead{(LAB)}  
& \colhead{(GB)}  
& \colhead{(GBT)}  
& \colhead{(Wak03)} 
& \colhead{ (HI4PI) } 
& \colhead{(3)--(4) }
& \colhead{(5)--(4) }
& \colhead{(6)--(4) }
 \\
   \colhead{(1)}
& \colhead{(2)}  
& \colhead{(3)}  
& \colhead{(4)}  
& \colhead{(5)}  
& \colhead{(6)}  
& \colhead{(7)}  
& \colhead{(8)}  
& \colhead{(9)}  }

\startdata

3C~249.1          & 20.400 & $20.431\pm0.013$ & $20.414\pm0.013$ & 20.446 (GB) & 20.452 & +0.017        & +0.032        & +0.038 \\
3C~273             & 20.169 & $20.184\pm0.021$ & $20.108\pm0.032$ & 20.222 (GB) & 20.231 & {\bf +0.076} & {\bf +0.114} & {\bf +0.123} \\
H1821+643       & 20.528 &          \dots              & $20.536\pm0.011$ & 20.584 (Eff)  & 20.549 & \dots            & +0.048         & +0.013         \\
HE~0226-4110  & 20.171 &          \dots              & $20.114\pm0.010$ & 20.272 (VE) & 20.165 &  \dots           &{\bf +0.158}   &{\bf +0.051} \\               
HS~0624+6907 & 20.791 &         \dots               & $20.781\pm0.024$ & 20.898 (Eff) & 20.802 &  \dots           &{\bf +0.117}   & +0.021  \\

MRC~2251-178 & 20.374 & $20.415\pm0.016$ & $20.390\pm0.014$ & 20.415 (GB)& 20.425 & +0.025         &  +0.025       & +0.035   \\
Mrk 205             & 20.436 &         \dots                & $20.463\pm0.012$ & 20.508 (Eff) & 20.468 & \dots            & +0.045        & +0.005  \\
Mrk 279             & 20.128 & $20.216\pm0.019$ & $20.193\pm0.020$  & 20.338 (Eff) & 20.104 & +0.023         & {\bf +0.145} & {\bf $-$0.089} \\                                                    
Mrk 335             & 20.496 & $20.555\pm0.015$ & $20.477\pm0.015$ & 20.567 (GB) & 20.517 & {\bf +0.078} & {\bf +0.090} & +0.040 \\              
Mrk 478             & 19.916 & $19.913\pm0.040$ & $19.927\pm0.035$ & 20.010 (GB) & 19.956 & $-$0.014      & {\bf +0.083} & +0.029 \\

Mrk 509             & 20.613 & $20.617\pm0.013$ & $20.583\pm0.012$ & 20.613 (GB)  & 20.594 & +0.034       &+0.030          & +0.011 \\ 
Mrk~771            & 20.415 & $20.357\pm0.016$ & $20.327\pm0.016$ & 20.377 (GB)  & 20.355 & +0.030       & {\bf +0.050}  & +0.028 \\  
Mrk 876             & 20.346 & $20.424\pm0.013$ & $20.395\pm0.015$ & 20.424 (Eff)   & 20.372 & +0.029       & +0.029         & $-$0.023  \\
Mrk~926            & 20.413 & $20.417\pm0.015$ & \dots                       & 20.436 (GB)  & 20.458 &  \dots          & \dots            & \dots     \\
Mrk 1383           & 20.387 & $20.382\pm0.017$ & $20.370\pm0.017$ & 20.547 (GB)  & 20.412 & +0.012        & {\bf +0.177} & +0.042  \\

Mrk 1513           & 20.550 & $20.544\pm0.013$ & $20.534\pm0.013$ & 20.555 (GB)  & 20.569 & +0.010        & +0.021        & +0.035  \\   
NGC~985          & 20.537 & $20.546\pm0.014$ & $20.534\pm0.015$ & 20.522 (GB) & 20.546 &  +0.012        & $-$0.012     & +0.012  \\                               
NGC~5548        & 20.131 & $20.172\pm0.022$ & $20.164\pm0.021$ & 20.203 (GB) & 20.190 &  +0.008        & +0.039         & +0.026  \\    
NGC~7469        & 20.654 & $20.643\pm0.021$ & $20.637\pm0.022$ & 20.649 (Eff)  & 20.654 & +0.006         & +0.012         & +0.017  \\
PG~0804+761   & 20.497 & $20.525\pm0.016$ & $20.514\pm0.017$ & 20.588 (Eff)  & 20.524 & +0.011         & {\bf +0.074} & +0.010  \\     
    
PG~0953+414   & 19.997 & $20.039\pm0.028$ & $19.995\pm0.020$ & 20.087 (Eff)  & 20.028 & +0.044         & {\bf +0.092} & +0.033 \\              
PG~1001+291   & 20.203 & $20.210\pm0.019$ & $20.195\pm0.019$ & 20.249 (GB) & 20.244 & +0.015         & {\bf +0.054} & +0.049  \\
PG~1116+215   & 20.065 & $20.084\pm0.035$ & $20.072\pm0.020$  & 20.075 (Eff)  & 20.086 & +0.012         & +0.003        & +0.014  \\ 
PG~1211+143   & 20.413 & $20.410\pm0.015$ & $20.413\pm0.013$ & 20.421 (GB) & 20.425 & $-$0.003       & +0.008        &  +0.012  \\
PG~1216+069   & 20.173 & $20.173\pm0.026$ & $20.163\pm0.021$ & 20.209 GB)  &20.182 &  +0.010          & +0.046        & +0.019  \\

PG~1259+593  & 19.983 & $20.168\pm0.037$ & $20.247\pm0.020$  & 20.225 (Eff) & 20.150 & {\bf $-$0.079} & $-$0.022    & {\bf $-$0.097}  \\        
PG~1302-102   & 20.497 & $20.485\pm0.014$ & $20.455\pm0.015$ & 20.502 (GB) & 20.494 &  +0.030         &  +0.047       & +0.039  \\
PG~1351+640  & 20.301 & $20.401\pm0.021$ & $20.419\pm0.022$ & 20.441 (Eff)  & 20.341 & $-$0.018       & +0.022        & {\bf $-$0.078}  \\
PG~1444+407  & 20.008 & $20.037\pm0.028$ & $20.022\pm0.029$ & 20.031 (GB)  & 20.036 &+0.015          & +0.009        & +0.014  \\
PHL~1811        & 20.606 & \dots                        & $20.599\pm0.019$ & 20.592 (LDS) & 20.628 & \dots           & $-$0.007      &  +0.029  \\    

PKS~0405-12   & 20.508 & $20.531\pm0.014$ & $20.509\pm0.013$ & 20.537 (GB) & 20.541 & +0.022          & +0.028          & +0.032 \\
PKS~2005-489 & 20.575 & \dots                       & \dots                       & 20.655 (VE)  & 20.561 & \dots             &  \dots             & \dots   \\
PKS~2155-304 & 20.117 & $20.092\pm0.026$ & $20.053\pm0.020$ & 20.123 (Eff)  & 20.102 & +0.039          & {\bf +0.070}   & +0.049  \\
Ton~S180         & 20.104 & $20.035\pm0.055$ & $20.031\pm0.028$ & 20.097 (GB) & 20.122 &+0.004           &  {\bf +0.066}  & {\bf +0.091} \\
Ton~S210         & 20.129 &    \dots                    & $20.137\pm0.010$  & 20.205 (Eff)  & 20.170 &  \dots           &{\bf +0.068}    & +0.033  \\
VII~Zw~118      & 20.543 & $20.564\pm0.015$ & $20.539\pm0.015$ & 20.558 (GB)  & 20.576 & +0.025         & +0.019           & +0.037  \\

\enddata

\tablenotetext{a} {Sample of 36 AGN sight lines surveyed in 21-cm emission at radio telescopes
 with different beam sizes.  Columns 2--4 list H~I column densities $\log N_{\rm HI}$ 
($N_{\rm HI}$ in cm$^{-2}$) from LAB, GB, GBT in Wakker \etal\ (2011).  Additional GBT data
in column~4 were provided by Jay Lockman (for Mrk~279, PG~1116+215, PG~1259+593, 
PKS~2155-304) and by Bart Wakker (for HE~0226-4110 and Ton~S210).  
Columns~5 and 6 list values for gas at all velocities (Wakker \etal\ 2003) 
and HI4PI survey. Telescope labels are:  
Leiden-Argentina-Bonn (LAB), Green Bank 140-ft (GB), 100-m Green Bank Telescope (GBT), 
Leiden-Dwingeloo Survey (LDS), Effelsberg (Eff), Villa Elisa (VE) with beam sizes:  LAB ($36'$), 
LDS ($35'$), VE ($34'$), GB ($21'$), HI4PI ($16'$), Eff ($9.7'$), GBT ($9.1'$).
Columns 7, 8, and 9 show differences $\Delta \log N_{\rm HI}$ relative to GBT in measurements 
from GB, Wakker \etal\ (2003), and HI4PI, respectively.  Values differing by more than 0.050 (dex) 
are highlighted in boldface.  The mean differences in $\log N_{\rm HI}$ from GBT values are:
 $+0.017$ (GB-140~ft), $+0.059$ (Wakker \etal\ 2003), and $+0.023$ (HI4PI).  
  }
    
\end{deluxetable}



\begin{deluxetable} {lcccc}
\tablecolumns{5}
 \tabletypesize{\footnotesize}

\tablenum{3}
\tablewidth{0pt}
\tablecaption{Nine AGN with Significant Molecular Fractions\tablenotemark{a}}  

\tablehead{
   \colhead{AGN Name}
& \colhead{$b$}  
& \colhead{$E(B-V)$ }
& \colhead{$\log N_{\rm H}$}  
& \colhead{$f_{\rm H2}$ }
\\
   \colhead{}
& \colhead{} 
& \colhead{(SF11 mag)} 
& \colhead{($N_{\rm H}$~cm$^{-2}$)} 
& \colhead{} 
}

\startdata
NGC~1068        & $-51.93$ & $0.0289\pm0.0004$  & 20.440 & 0.0098  \\
PG~1351+640   & $+52.02$ & $0.0177\pm0.0003$  & 20.448 & 0.0156  \\
PG~1211+143   & $+74.32$ & $0.0286\pm0.0011$  & 20.429 & 0.0179   \\
Mrk~335            & $-41.42$ & $0.0305\pm0.0031$  & 20.583 & 0.0353   \\
NGC~7714        & $-55.56$ & $0.0451\pm0.0002$  & 20.690 & 0.0356  \\
Mrk~116            & $+44.84$ & $0.0292\pm0.0028$  & 20.537 & 0.0715  \\
PHL~1811         & $-44.81$ & $0.0415\pm0.0031$  & 20.640 & 0.105   \\
NGC~3310        & $+54.06$ & $0.0192\pm0.0005$  & 20.227 & 0.164   \\   
NGC~7469        & $-45.47$ & $0.0599\pm0.0015$  & 20.732 & 0.173  \\
\enddata 

\tablenotetext{a} {Nine high-latitude AGN ($|b| \geq 40\degr$) from group~1 
with molecular fractions greater than 1\%.  Galactic latitudes ($b$), $E(B-V)$, 
and total hydrogen column densities $N_{\rm H}$ are from Table 1.  
Molecular fractions $f_{\rm H2} \equiv 2N_{\rm H2}/N_{\rm H}$ often
exceed 1\% when $\log N_{\rm H} > 20.38\pm0.13$, the 
atomic-to-molecular transition seen toward high-latitude  AGN
(Gillmon \etal\ 2006).
  }
    
\end{deluxetable}



\begin{deluxetable} {llccl}
\tablecolumns{5}

\tabletypesize{\footnotesize}

\tablenum{4}
\tablewidth{0pt}
\tablecaption{Dust Emission Model Parameters\tablenotemark{a} }  

\tablehead{
   \colhead{Paper}
& \colhead{Data~Set}
& \colhead{$T_d$~(K) }
& \colhead{Index $\beta$ }  
& \colhead{Comments}  
 }

\startdata
SFD98                & IRAS/DIRBE    & 18.2         &  2.0             &  Calibrated on elliptical galaxies \\
SF11                   & Recalibration   & 18.2         &  2.0             &  Calibrated on SDSS stars   \\
Planck 2011        & Planck-HIFI     & 17.9(0.9)  & 1.78(0.18)   &  Anticorrelation $\beta \propto T_d^{-2/3}$  \\
Planck 2014        & Planck-HIFI     & 20.3(1.3) & 1.59(0.12)   &  Values at $|b| > 15^{\circ}$   \\
Planck 2016        & Planck-HIFI     & 19.4(1.5) & 1.63(0.17)   &  Values at  $|b| > 20^{\circ}$ (353~GHz) \\        
Planck 2020        & Planck HIFI      & 19.6        & 1.55(0.05)   &  Polarized dust emission  \\
Casandjian 2022 & IRAS/Planck    & 20.2        & 1.4               & Excess dust at low $N_{\rm HI}$  
\enddata 

\tablenotetext{a} {Mean values of dust temperature $T_d$, with dispersion ($1\sigma$) in parentheses,
and emissivity index ($\beta$) measured or assumed in a sample of FIR papers:  
Schlegel \etal\ (1998), Schlafly \& Finkbeiner (2011), Casandjian \etal\ (2022), and four papers
from the Planck Collaboration (listed in references).  Meisner \& Finkbeiner (2015) explored a 
two-temperature model with ``hot dust" ($T_2 \approx 16.2$~K, $\beta_2 = 1.67$) and 
``cold dust" ($T_1 \approx 9.15$~K, $\beta_1 = 2.70$). }
  
\end{deluxetable}



\begin{deluxetable} {lllcc}
\tablecolumns{5}

\tabletypesize{\scriptsize}

\tablenum{5}
\tablewidth{0pt}
\tablecaption{Far-IR, E(B-V), and Dust Temperatures\tablenotemark{a} }  

\tablehead{
\colhead{ AGN Name } 
& \colhead{ $I(100~\mu$m) }
& \colhead{ $E(B-V)$ }
& \colhead{ $I(100)/E(B-V)$ }  
& \colhead{ $T_{\rm d}$ }  
\\
\colhead{ } 
& \colhead{ (MJy/sr) }
& \colhead{(mag)}
& \colhead{(MJy/sr/mag)}  
& \colhead{ (K) }  
}

\startdata

3C~249.1          &  $1.66\pm0.11$ & $0.0301\pm0.0020$ & $55.0\pm5.2$ & $17.861\pm0.006$   \\
3C~273             & $1.04\pm0.03$ & $0.0179\pm0.0004$ & $58.1\pm2.1$ & $17.978\pm0.013$   \\     
ESO~141-G55  & $5.09\pm0.30$ & $0.0944\pm0.0062$ & $53.9\pm4.8$ & $17.861\pm0.006$  \\   
Fairall~9            & $1.13\pm0.06$ & $0.0217\pm0.0011$ & $52.1\pm3.7$ & $17.740\pm0.011$  \\
H1821+643       & $2.14\pm0.03$ & $0.0370\pm0.0007$ & $57.8\pm1.4$ & $17.974\pm0.019$  \\    
HE~0226-4110  & $0.75\pm0.03$ & $0.0132\pm0.0005$ & $56.7\pm3.1$ & $17.939\pm0.014$ \\                    
HE~1143-1810  & $1.82\pm0.03$ & $0.0331\pm0.0005$ & $55.0\pm1.2$ & $17.856\pm0.010$ \\  
HS~0624+6907 & $3.53\pm0.08$ & $0.0845\pm0.0021$ & $41.8\pm1.4$ & $17.278\pm0.017$  \\   

MRC~2251-178 & $1.87\pm0.06$ &$0.0335\pm0.0011$ & $55.8\pm2.6$  & $17.893\pm0.004$ \\   
Mrk~9                &  $2.40\pm0.08$ &$0.0503\pm0.0015$ & $47.7\pm2.1$ & $17.550\pm0.010$  \\   
Mrk 106             &  $1.25\pm0.05$ &$0.0235\pm0.0010$ & $53.2\pm3.1$ & $17.780\pm0.009$  \\                              
Mrk 116             & $1.68\pm0.16$ & $0.0292\pm0.0028$ & $57.5\pm7.8$ & $17.966\pm0.006$  \\                
Mrk 205             & $1.82\pm0.06$ & $0.0344\pm0.0011$ & $52.9\pm2.4$ & $17.781\pm0.009$   \\                           
Mrk 209             & $0.68\pm0.03$ & $0.0122\pm0.0005$ & $55.3\pm3.3$  & $17.876\pm0.005$  \\    
Mrk 279             & $0.80\pm0.03$ & $0.0138\pm0.0005$ & $57.9\pm3.0$  & $17.969\pm0.005$  \\                                        
Mrk 290             & $0.66\pm0.05$ & $0.0120\pm0.0008$ &$ 55.0\pm5.6$ & $17.871\pm0.004$    \\          
Mrk 335             & $1.58\pm0.16$ & $0.0305\pm0.0031$ & $51.8\pm7.4$ & $17.734\pm0.013$   \\                          
Mrk 421             & $0.73\pm0.06$ & $0.0130\pm0.0011$ & $56.0\pm6.6$ & $17.893\pm0.003$   \\          
Mrk 478             &  $0.64\pm0.06$ & $0.0111\pm0.0009$ & $57.3\pm7.1$ & $17.945\pm0.005$   \\     
Mrk 501             &  $0.87\pm0.02$ & $0.0164\pm0.0003$ & $53.0\pm1.6$ & $17.782\pm0.015$  \\     
Mrk 509             &  $2.76\pm0.03$ & $0.0492\pm0.0005$ & $56.0\pm0.8$ &  $17.905\pm0.005$  \\     
Mrk 817             &  $0.34\pm0.01$ & $0.0059\pm0.0002$ & $57.2\pm2.6$ & $17.948\pm0.003$   \\                        
Mrk 876             &  $1.38\pm0.06$ & $0.0230\pm0.0009$ & $59.8\pm2.3$ & $18.049\pm0.006$   \\     
Mrk 1095           & $5.61\pm0.11$ & $0.1099\pm0.0017$  & $51.1\pm1.3$ & $17.700\pm0.005$  \\     
Mrk 1383           & $1.78\pm0.06$ & $0.0276\pm0.0009$ & $64.5\pm3.0$ & $18.216\pm0.009$  \\     
Mrk 1513           & $2.07\pm0.09$ & $0.0370\pm0.0017$ & $55.9\pm3.8$ & $17.900\pm0.013$  \\    
MS~0700.7+6338 &$1.87\pm0.04$ & $0.0448\pm0.0014$ & $41.8\pm1.6$ & $17.273\pm0.020$  \\     
                    
NGC~985          & $1.60\pm0.02$ & $0.0282\pm0.0003$ & $56.7\pm0.9$ & $17.931\pm0.019$   \\                                       
NGC~1068        & $2.28\pm0.02$ & $0.0289\pm0.0004$ & $78.7\pm1.3$ & $18.709\pm0.017$  \\       
NGC~1399        & $0.60\pm0.02$ & $0.0109\pm0.0004$ & $54.8\pm2.7$ & $17.541\pm0.004$ \\                                           
NGC~1705        & $0.40\pm0.03$ & $0.0070\pm0.0005$ & $57.8\pm6.0$ & $17.980\pm0.002$ \\     
NGC~3310        & $1.12\pm0.03$ & $0.0192\pm0.0005$ & $58.3\pm2.2$ & $17.992\pm0.010$ \\            
NGC~3690        & $0.83\pm0.02$ & $0.0144\pm0.0002$ & $57.7\pm1.6$ & $17.966\pm0.010$ \\        
NGC~4151        & $1.45\pm0.07$ & $0.0237\pm0.0011$ & $61.3\pm4.1$ & $18.103\pm0.010$ \\    
NGC~4214        & $1.17\pm0.02$ & $0.0187\pm0.0003$ & $62.8\pm1.5$ & $18.154\pm0.014$ \\        
NGC~4670        & $0.75\pm0.02$ & $0.0128\pm0.0003$ & $58.6\pm2.1$ & $18.013\pm0.006$  \\                      
NGC~7469        & $2.92\pm0.07$ & $0.0599\pm0.0015$ & $48.7\pm1.7$ & $17.598\pm0.024$  \\     
NGC~7714        & $2.19\pm0.01$ & $0.0451\pm0.0002$ & $48.5\pm0.3$ & $17.588\pm0.005$  \\       

PG~0804+761   & $1.61\pm0.03$ & $0.0302\pm0.0005$ & $53.3\pm1.3$ & $17.793\pm0.004$ \\              
PG~0844+349   & $1.72\pm0.05$ & $0.0314\pm0.0007$ & $54.6\pm2.0$ & $17.848\pm0.019$ \\     
PG~0953+414   & $0.57\pm0.03$ & $0.0102\pm0.0006$ & $55.7\pm4.4$ & $17.898\pm0.001$ \\                        
PG~1116+215   & $1.13\pm0.01$ & $0.0193\pm0.0001$ & $58.5\pm0.6$ & $18.001\pm0.003$ \\          
PG~1211+143   & $1.76\pm0.06$ & $0.0286\pm0.0011$ & $61.6\pm3.2$ & $18.118\pm0.022$ \\     
PG~1259+593   & $0.41\pm0.02$ & $0.0070\pm0.0004$ & $58.6\pm4.4$ & $17.995\pm0.003$ \\               
PG~1302-102   & $2.38\pm0.06$ & $0.0376\pm0.0009$ & $63.2\pm2.2$  & $18.172\pm0.003$ \\            
PG~1351+640  & $1.15\pm0.03$ & $0.0177\pm0.0003$ & $65.0\pm2.0$  & $18.237\pm0.016$ \\       
PG~1626+554  & $0.29\pm0.03$ & $0.0053\pm0.0005$ & $55.3\pm7.7$ & $17.877\pm0.012$ \\
  
PHL~1811         & $2.15\pm0.15$ & $0.0415\pm0.0031$ & $51.8\pm5.3$ & $17.733\pm0.012$ \\               
PKS~0405-12   & $2.74\pm0.20$ & $0.0503\pm0.0036$ & $54.6\pm5.6$ & $17.845\pm0.008$ \\       
PKS~0558-504 & $2.37\pm0.07$ & $0.0388\pm0.0012$ & $61.2\pm2.6$ & $18.105\pm0.007$  \\         
PKS~2005-489 & $2.94\pm0.03$ & $0.0479\pm0.0007$ & $61.4\pm1.1$ & $18.108\pm0.012$ \\      
PKS~2155-304 & $1.10\pm0.03$ & $0.0185\pm0.0005$ & $59.3\pm2.3$ & $18.036\pm0.010$ \\      
Ton~S180         & $0.72\pm0.02$ & $0.0123\pm0.0002$ & $58.1\pm1.9$ & $17.982\pm0.002$ \\      
Ton~S210         & $0.84\pm0.03$ & $0.0144\pm0.0004$ & $58.2\pm2.6$ & $17.987\pm0.003$ \\      
VII~Zw~118      & $1.62\pm0.04$ & $0.0330\pm0.0008$ & $49.2\pm1.7$ & $17.624\pm0.008$   \\            
                           &                        &                                  &                        &                       \\
ESO350-IG38   & $0.55\pm0.01$ & $0.0096\pm0.0001$ & $57.8\pm1.0$ & $17.969\pm0.001$   \\      
ESO572-G34    & $1.87\pm0.03$ & $0.0340\pm0.0004$ & $54.9\pm1.0$ & $17.862\pm0.009$  \\      
HE~0238-1904  & $1.54\pm0.06$ & $0.0272\pm0.0011$ & $56.5\pm3.2$ & $17.922\pm0.063$   \\
Mrk 487             & $0.67\pm0.02$ & $0.0120\pm0.0003$ & $55.8\pm2.1$ & $17.894\pm0.006$   \\
NGC~4649        & $1.38\pm0.04$ & $0.0226\pm0.0006$ & $61.0\pm2.4$ & $18.094\pm0.012$   \\
HE~1228+0131 & $0.95\pm0.08$ & $0.0162\pm0.0014$ & $54.5\pm6.6$ & $18.017\pm0.010$    \\
HE~1326-0516  & $1.64\pm0.04$ & $0.0256\pm0.0006$ & $64.1\pm2.2$ & $18.213\pm0.006$   \\
HS~1102+3441 & $1.11\pm0.07$ & $0.0197\pm0.0012$ & $56.5\pm4.8$ & $17.925\pm0.004$   \\
Mrk~36             & $1.48\pm0.07$ & $0.0250\pm0.0012$ & $59.1\pm4.0$ & $18.028\pm0.002$   \\
Mrk~54             & $0.73\pm0.02$ & $0.0129\pm0.0004$ & $56.7\pm2.3$ & $17.928\pm0.025$   \\

Mrk~59             & $0.53\pm0.02$ & $0.0093\pm0.0004$ & $57.3\pm3.5$ & $17.950\pm0.005$   \\
Mrk~487           & $0.67\pm0.02$ & $0.0120\pm0.0003$ & $55.8\pm2.1$ & $17.894\pm0.006$   \\
Mrk~734           & $1.49\pm0.06$ & $0.0267\pm0.0011$ & $55.8\pm3.2$ & $17.892\pm0.004$   \\
Mrk~771           & $1.24\pm0.02$ & $0.0233\pm0.0003$ & $53.3\pm1.0$ & $17.794\pm0.007$   \\
Mrk~829           & $0.61\pm0.01$ & $0.0108\pm0.0003$ & $56.2\pm2.0$ & $17.917\pm0.004$   \\
Mrk~926           & $2.07\pm0.04$ & $0.0354\pm0.0007$ & $58.5\pm1.6$ & $17.997\pm0.006$   \\
Mrk~1502         & $2.55\pm0.08$ & $0.0559\pm0.0018$ & $45.6\pm2.0$ & $17.455\pm0.003$   \\
NGC~3504       & $1.32\pm0.01$ & $0.0228\pm0.0002$ & $58.1\pm0.7$ & $17.991\pm0.008$   \\
NGC~3991       & $1.06\pm0.01$ & $0.0189\pm0.0002$ & $56.1\pm0.9$ & $17.901\pm0.004$   \\ 

NGC~5548       & $0.92\pm0.05$ & $0.0168\pm0.0009$ & $54.5\pm4.1$ & $17.838\pm0.005$   \\ 
NGC~7496       & $0.47\pm0.01$ & $0.0084\pm0.0001$ & $55.9\pm1.0$ & $17.887\pm0.003$   \\     
PG~0947+396  & $0.92\pm0.04$ & $0.0162\pm0.0006$ & $56.9\pm3.1$ & $17.943\pm0.002$   \\
PG~1001+291  & $1.08\pm0.03$ & $0.0190\pm0.0005$ & $57.1\pm2.0$ & $17.924\pm0.002$   \\
PG~1004+130  & $1.87\pm0.02$ & $0.0331\pm0.0005$ & $56.6\pm1.1$ & $17.928\pm0.003$   \\
PG~1048+342  & $1.13\pm0.07$ & $0.0199\pm0.0012$ & $56.8\pm5.0$ & $17.933\pm0.002$    \\
PG~1216+069  & $1.01\pm0.04$ & $0.0186\pm0.0007$ & $54.4\pm2.9$ & $17.845\pm0.005$    \\
PG~1307+085  & $1.58\pm0.05$ & $0.0292\pm0.0008$ & $54.0\pm2.2$ & $17.823\pm0.005$    \\
PG~1352+183  & $0.90\pm0.04$ & $0.0159\pm0.0006$ & $56.7\pm3.1$ & $17.931\pm0.011$    \\
PG~1402+261  & $0.74\pm0.02$ & $0.0132\pm0.0004$ & $56.1\pm2.4$ & $17.907\pm0.008$    \\

PG~1404+226  & $1.12\pm0.02$ & $0.0190\pm0.0003$ & $58.7\pm1.3$ & $18.007\pm0.002$   \\
PG~1411+442  & $0.48\pm0.07$ & $0.0084\pm0.0012$ & $56.5\pm11.7$ & $17.927\pm0.014$  \\
PG~1415+451  & $0.40\pm0.02$ & $0.0071\pm0.0003$ & $56.9\pm3.4$ & $17.928\pm0.009$   \\ 
PG~1444+407  & $0.63\pm0.02$ & $0.0114\pm0.0004$ & $55.0\pm2.8$ & $17.855\pm0.005$   \\
PG~2349-014   & $1.33\pm0.02$ & $0.0234\pm0.0004$ & $56.9\pm1.3$ & $17.921\pm0.009$   \\
SBS~0335-052 & $2.05\pm0.07$ & $0.0402\pm0.0015$ & $51.0\pm2.7$ & $17.698\pm0.012$   \\
SBS~1415+437 & $0.45\pm0.02$ & $0.0077\pm0.0003$ & $58.6\pm3.4$ & $18.007\pm0.003$  \\
Tol~0440-381    & $0.67\pm0.03$ & $0.0130\pm0.0005$ & $51.2\pm2.9$ & $17.711\pm0.012$   \\
Ton~1187          & $0.52\pm0.04$ & $0.0093\pm0.0008$ & $56.0\pm6.7$ & $17.894\pm0.002$   \\
vZ~1128            & $0.66\pm0.03$ & $0.0117\pm0.0005$ & $56.1\pm3.3$ & $17.909\pm0.002$  \\

\enddata 

\tablenotetext{a} {FIR data (SF11) used to estimate the extinction toward 94 AGN in Table 1.   
The first 55 quasars have both H~I and H$_2$ measurements;  the next 39 have only H~I.  
Columns 2 and 3 are the 100~$\mu$m surface brightness and FIR-inferred color excess,
tabulated on the IPAC/IRSA website ({\url {https://irsa.ipac.caltech.edu/applications/DUST}}).  
Column~4 gives the ratio, $I(100~\mu$m)/$E(B-V)$, with errors found from the relative errors 
on $I(100~\mu$m) and $E(B-V)$ added in quadrature.  Column~5 gives the modeled dust 
temperature $T_d$.  Several AGN sight lines have large errors on $I(100~\mu$m) and $E(B-V)$ 
with their ratios uncertain by 10-15\%.  For PG~1411+442, the uncertainty is 20\%.  }
  
\end{deluxetable}



\begin{deluxetable} {lccccc}
\tablecolumns{6}

\tabletypesize{\scriptsize}

\tablenum{6}
\tablewidth{0pt}
\tablecaption{Statistics of QSO Sub-samples\tablenotemark{a} }  

\tablehead{
   \colhead{ Sample }
& \colhead{ $N_{\rm QSO}$ }
& \colhead{$\langle E(B-V) \rangle$ }
& \colhead{ $\langle N_{\rm H}/E(B-V) \rangle$ }  
& \colhead{ $\langle N_{\rm H}/E(B-V) \rangle$ }  
& \colhead{ $\langle N_{\rm H}/E(B-V) \rangle$ }  
 \\
  \colhead{}
& \colhead{}  
& \colhead{(SF11)}  
& \colhead{(all vel/SF11)}  
& \colhead{(low vel/SF11)}  
& \colhead{(all vel/Pl-GN)}  
 \\
   \colhead{(1)}
& \colhead{(2)}  
& \colhead{(3)}  
& \colhead{(4)}  
& \colhead{(5)}  
& \colhead{(6)}  
}

\startdata
Primary (H\,I, H$_2$) & 55 & 0.0295 & $11.53\pm0.57$  & $9.66\pm0.34$ & $10.59\pm0.42$   \\ 
Modified                     &  51 & 0.0313 & $10.64\pm0.41$ & $9.22\pm0.27$ & $10.30\pm0.38$    \\
Complex~C                &   9 & 0.0151 & $17.0\pm5.2$      & $10.7\pm3.5$   & $12.9\pm4.3$        \\  
Low-E(B-V)                &   4 & 0.0063 &  22.9                    & 15.4                  & 14.3                       \\                         
                                   &       &             &                            &                          &                                \\
Secondary (H\,I)         &  39 & 0.0192 & $10.20\pm0.38$ & $9.37\pm0.38$ & $9.27\pm0.35$    \\
Modified                     &   32 & 0.0219 & $9.66\pm0.34$   & $9.03\pm0.37$ & $9.02\pm0.27$      \\
Low-E(B-V)                &    7  & 0.0084 & 12.7                    & 10.9                  & 10.4                       \\
                              
\enddata 

\tablenotetext{a} {Mean values of $E(B-V)$ and gas-to-dust ratios 
$\langle N_{\rm H}/E(B-V) \rangle$ in units 
$10^{21}~{\rm cm}^{-2}~{\rm mag}^{-1}$ for various sub-samples of the 94 quasars. 
The mean ratios are shown for gas at all velocities (columns 4 and 6) and low velocity
only ($|V_{\rm LSR}| \leq 90$~\kms) in column 5.   Errors on the mean are 
evaluated as $\sigma / N^{1/2}$, where $\sigma$ is the variance in sample of $N$ targets.  
Ratios in columns 4 and 5 adopt $E(B-V)$ from SF11, while those in column~6 
adopt Planck-GN.  The primary group includes 55 QSOs with both H~I 
and H$_2$ measurements.  In the secondary group of 39 QSOs with only H~I, 
the mean ratios are 12\% lower, reflecting the missing contribution from H$_2$.  
In each group, the modified samples exclude sight lines with low (and uncertain) 
$E(B-V) < 0.01$ from SF11. Nine quasars behind HVC Complex~C (Table~7) are 
among the 55 targets with both H~I and H$_2$.  Their elevated ratios exhibit large 
variations in FIR estimates of $E(B-V)$ from SF11 and Planck-GN.}

\end{deluxetable}


\begin{deluxetable} {lccc cccc}
\tablecolumns{8}

\tabletypesize{\scriptsize}

\tablenum{7}
\tablewidth{0pt}
\tablecaption{Ratios\tablenotemark{a} in High-Velocity Cloud Complex-C }  

\tablehead{
   \colhead{QSO}
& \colhead{$E(B-V)$ }
& \colhead{$E(B-V)$ }
& \colhead{$E$-Ratio}
& \colhead{Ratio-all}  
& \colhead{Ratio-low}  
& \colhead{Ratio-all}  
& \colhead{Ratio-low}  
\\
   \colhead{  }
& \colhead{(SF11)} 
& \colhead{(Pl-GN)} 
& \colhead{(GN/SF11)} 
& \colhead{(SF11)} 
& \colhead{(SF11)} 
& \colhead{(Pl-GN)} 
& \colhead{(Pl-GN)} 
 \\
   \colhead{(1)}
& \colhead{(2)}  
& \colhead{(3)}  
& \colhead{(4)}  
& \colhead{(5)}  
& \colhead{(6)}  
& \colhead{(7)}  
& \colhead{(8)}  }

\startdata
Mrk~205          & $0.0344\pm0.0011$ & $0.0377\pm0.0053$ & 1.10 & 9.37  & 7.38 & 8.54 & 6.73   \\ 
Mrk~279          & $0.0138\pm0.0005$ & $0.0154\pm0.0002$ & 1.12 & 15.8  & 6.11 & 14.2 & 5.49  \\
Mrk~290          & $0.0120\pm0.0008$ & $0.0162\pm0.0003$ & 1.35 & 22.2  & 11.1 & 16.4 & 8.21  \\ 
Mrk~501          & $0.0164\pm0.0003$ & $0.0183\pm0.0004$ & 1.11 & 12.6  & 10.6 & 11.3 & 9.54  \\ 
Mrk~817          & $0.0059\pm0.0002$ & $0.0115\pm0.0001$ & 1.95 & 20.6   & 15.1 &10.6 & 7.75  \\ 
Mrk~876          & $0.0230\pm0.0009$ & $0.0291\pm0.0006$ & 1.26 & 11.5  & 7.05 & 9.14 & 5.57  \\
PG~1259+593 & $0.0070\pm0.0004$ & $0.0086\pm0.0001$ & 1.23 & 24.0  & 11.2 & 19.5 & 5.43  \\ 
PG~1351+640 & $0.0177\pm0.0003$ & $0.0156\pm0.0001$ & 0.88 & 15.9  & 11.2 & 18.0 & 12.7  \\ 
PG~1626+554 & $0.0053\pm0.0005$ & $0.0137\pm0.0002$ & 2.59 & 21.3  & 16.3 & 8.24 & 6.28   \\        
                         &                                  &                                  &           &          &         &         &          \\            
Mean values     & 0.0151                      & 0.0185                      &           & 17.0  & 10.7 & 12.9 & 7.5    \\ 
\enddata 

\tablenotetext{a} {Reddening values $E(B-V)$ and gas-to-dust ratios $N_{\rm H}/E(B-V)$ 
(in $10^{21}~{\rm cm}^{-2}~{\rm mag}^{-1}$) for gas at all velocities and low velocity only
toward nine sight lines to quasars behind HVC Complex C (Figure 2). 
Columns 2 and 3 list $E(B-V)$ from SF11 and Planck-GN (Planck Collaboration XLVIII 2016), 
with their ratio in column (4).  Columns 5 and 6 list gas-to-dust ratios $N_{\rm H}/E(B-V)$ based 
on SF11 reddening and gas at different velocities. Columns 7 and 8 list the ratios based 
on Planck-GN reddening.  Mean values are listed at the bottom, spanning a range outside 
the errors.  Mean values of $E(B-V)$ from Planck-GN are 22\% higher than SF11, with
most $E(B-V)$ ratios higher than SF11 values.  Two sight lines have GN reddening 
1.9--2.6 times larger than SF11, and only one of nine has a ratio less than 1.0.  The mean 
low-velocity ($|V_{\rm LSR}| < 90$ km~s$^{-1}$) gas-to-dust ratio is $\sim60$\% of that 
for gas at all velocities.}
    
\end{deluxetable}

\clearpage


\begin{figure}[ht]
\includegraphics[angle=0,scale=0.80]{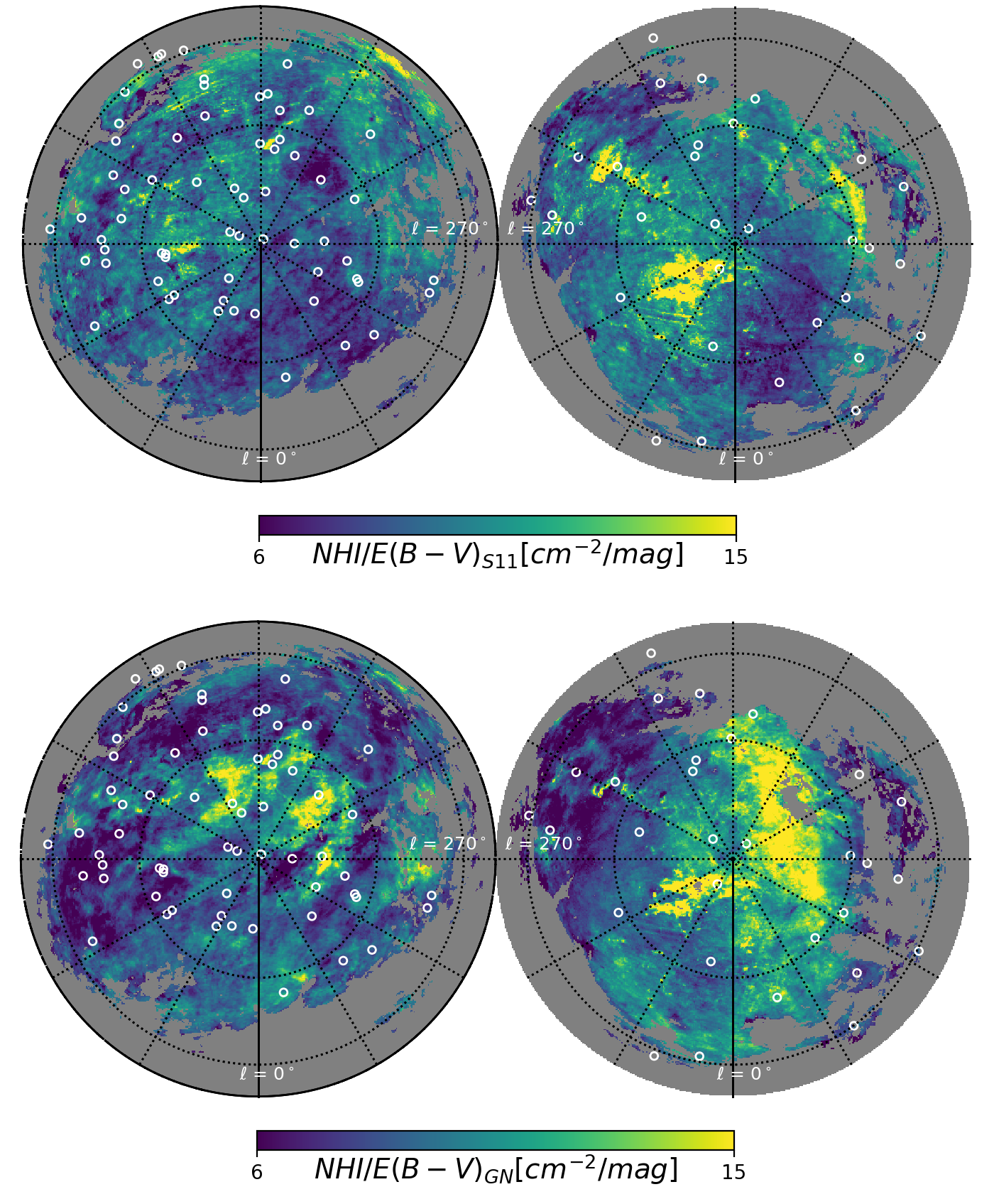}
\caption{\small {Polar projection maps of the ratio \Ratio\ between $N_{\rm HI}$ from HI4PI 
for low-velocity gas with $|V_{\rm LSR}| \leq 90$~\kms\ and $E(B-V)$ from SF11 (top) and 
Planck-GN (bottom).  Locations of the 94 quasars are shown as white circles.
Both reddening maps are smoothed to the HI4PI resolution of $16'$.   
Northern Galactic hemisphere is on the left, southern hemisphere on the right.  
Galactic longitude $\ell = 0\degr$ is at the bottom and $\ell = 180\degr$ at the top of
each map.  Longitude increases clockwise for northern hemisphere and counterclockwise for 
southern hemisphere, meeting at $\ell = 270\degr$.  Color scale at bottom shows 
gas-to-dust ratio in units $10^{21}~{\rm cm}^{-2}~{\rm mag}^{-1}$. The linear scale covers
the $1^{\rm st}$ and $99^{\rm th}$ percentiles of the distribution of ratios of the top panel. 
White circles mark locations of 94 AGN.}
 }
\end{figure}
 


\begin{figure}[ht]
\includegraphics[angle=0,scale=0.53]{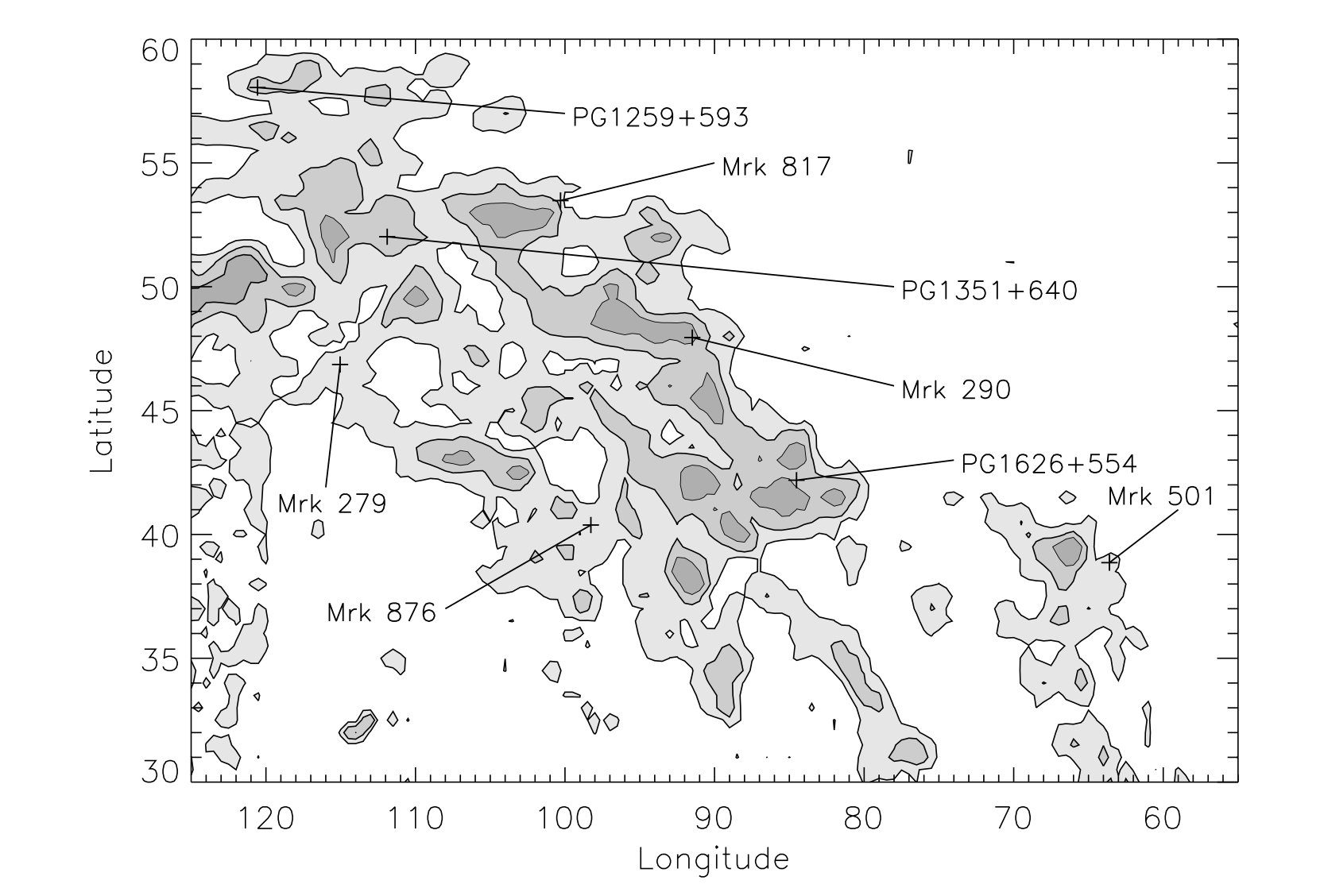}
\caption{Contours of H~I column density in Complex~C from the Leiden-Dwingeloo Survey
(Hartmann \& Burton 1997) plotted on a $0.5\deg$ grid ($35'$ beam) with H~I contours at 
1, 3, and $6 \times 10^{19}~{\rm cm}^{-2}$.   Locations of 8 of the 9 quasars in our study 
(Table 5) are labeled;  Mrk~205 ($\ell = 125.45^{\circ}$, $b = 41.67^{\circ}$) lies just off the
left side of the plot.  Complex~C extends over $\sim2000$ deg$^2$ in the Northern Galactic 
sky, with metallicity along these sight lines ranging from 0.1--0.3 solar 
(Collins \etal\ 2003, 2007).} 
\end{figure}
 


\begin{figure}[ht]
\includegraphics[angle=0,scale=1.3]{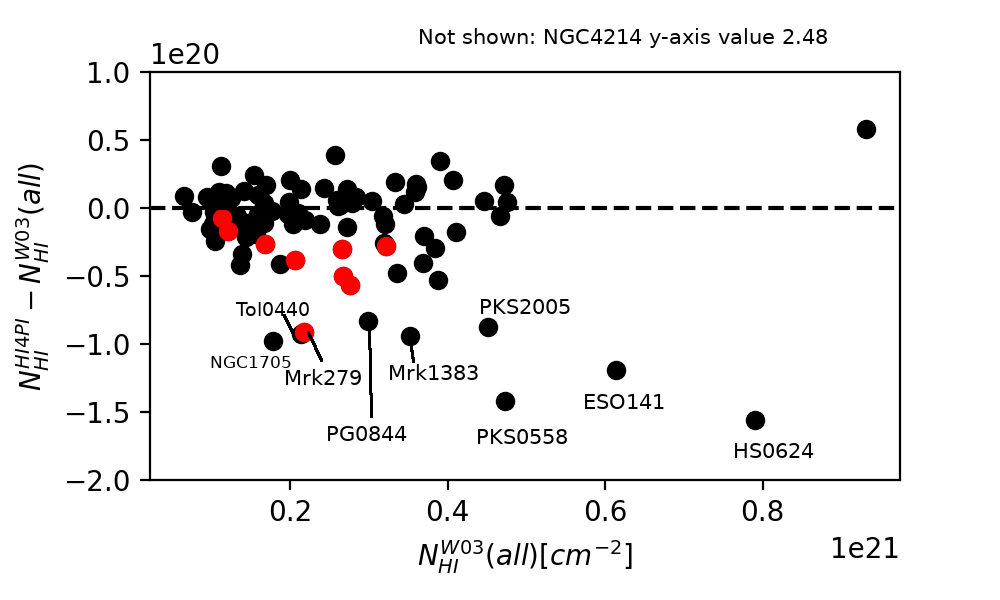}
\caption{Differences in $N_{\rm HI}$ (at all velocities, in units of 
$10^{20}~{\rm cm}^{-2}$) between HI4PI values (HI4PI Collaboration \etal\ 2016) 
 and those from Wakker \etal\ (2003) from our Table 1.  These are plotted
 versus $N_{\rm HI}$ (in units of $10^{21}~{\rm cm}^{-2}$) from Table 1.  
 Red points indicate nine sight lines passing through Complex~C, an HVC with
 low dust content where \HI\ makes a significant contribution to total $N_{\rm HI}$.   
 We have annotated several sight lines in which $N_{\rm HI}$ from HI4PI is 
 considerably lower than that in Wakker \etal\ (2003).  NGC~4214 is not shown, 
 because its internal \HI\ (at $+295$~\kms) is included in the HI4PI velocity range.   
}
 
\end{figure}
 


\begin{figure}[h]
\includegraphics[angle=0,scale=1.35]{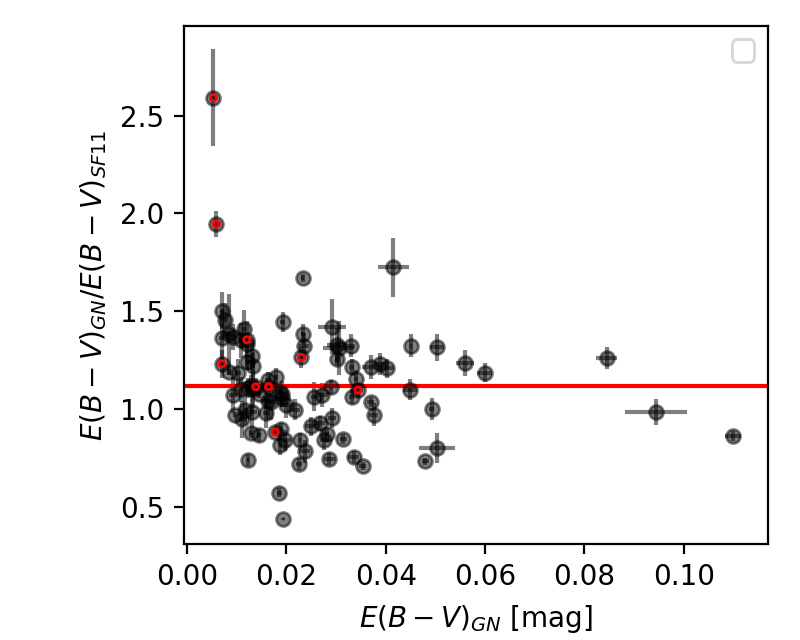}
\caption{Distribution of ratios of $E(B-V)$ from Planck-GN and SF11.  Values are 
from full sample (94 AGN), with nine QSOs behind HVC Complex~C plotted in 
red.  These include outliers (PG~1626+554 and Mrk~817) with anomalously 
high GN/SF11 ratios and very low $E(B-V)$.  On average, the Planck-GN values 
are 12\% higher than SF11, with considerable dispersion about the unweighted 
mean (red line), particularly in sight lines with $E(B-V) \lesssim 0.04$. The plotted 
errors are likely under-estimated owing to systematic effects in FIR modeling.   }
\end{figure}
 


\begin{figure}[ht]
\includegraphics[angle=0,scale=1.1]{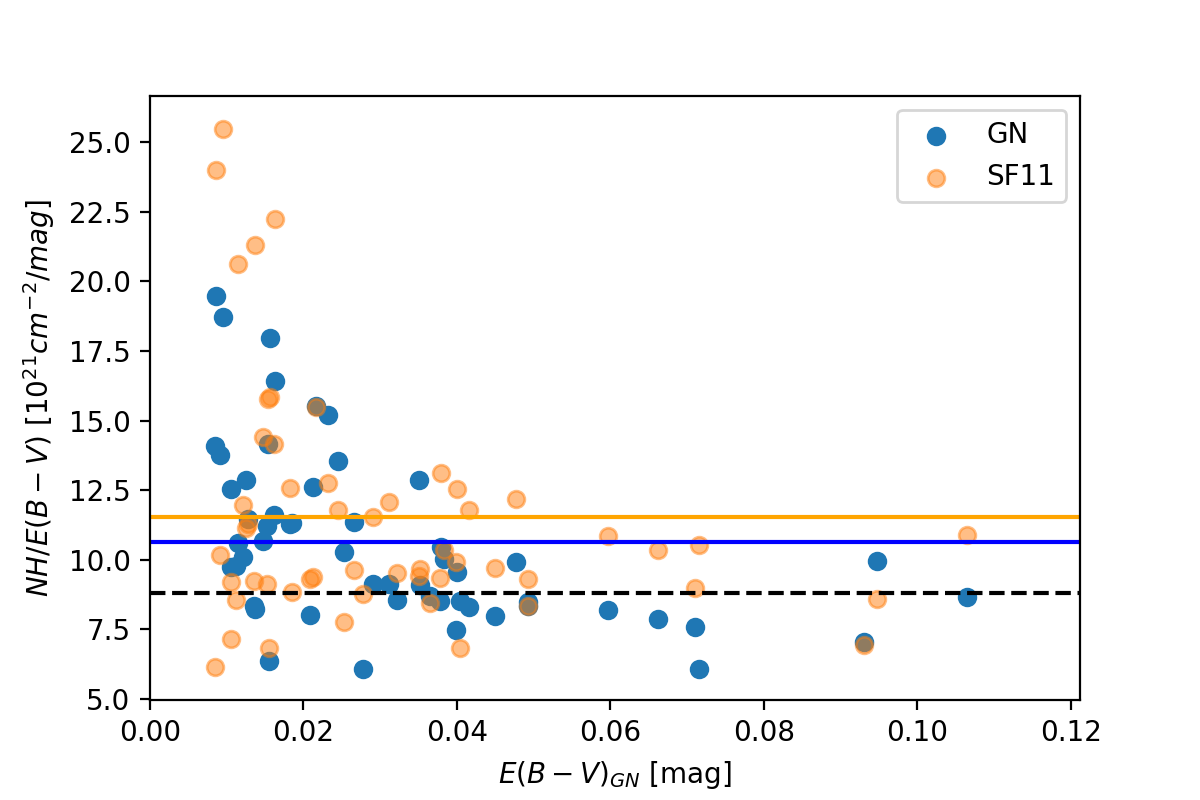}
\caption{Gas-to-dust ratios (in $10^{21}~{\rm cm}^{-2}~{\rm mag}^{-1}$) using total 
hydrogen column density ($N_{\rm H}$) for 55 AGN in our primary sample with both
\HI\ and \Htwo\ (Table 1).  The FIR-inferred $E(B-V)$ are plotted for both Planck-GN 
(blue circles) and SF11 (orange circles).  Horizontal lines show the mean ratios of 
two distributions:  10.6 (blue solid line, Planck-GN), 11.5 (orange solid line, SF11). 
Black dotted line shows the mean ratio $N_{\rm HI}/E(B-V) = 8.8$ from
Lenz \etal\ (2017), which was based on just \HI\ column densities. }
\end{figure}
 


\begin{figure}[ht]
\includegraphics[angle=0,scale=0.9] {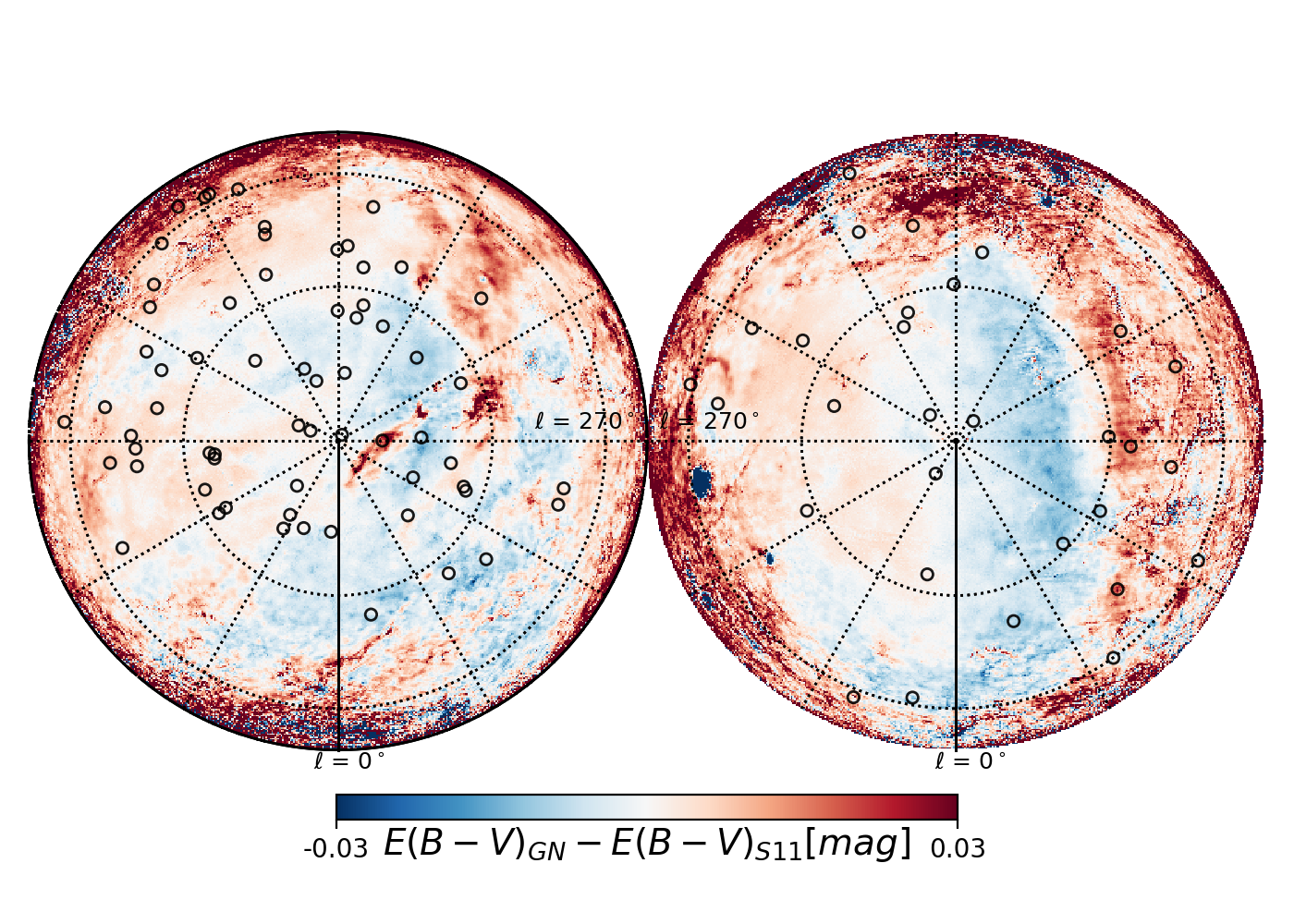}
\caption{ Polar projection maps showing the difference between $E(B-V)$ 
from Planck-GN and values from SF11.  These variations track changes in the 
gas-to-dust ratio seen in the two ratio maps of Figure 1. For example, the red 
feature towards ($\ell, b) = (260-330\degr,  80-84\degr)$ is Markkanen's Cloud
(Markkanen 1979) with a discrepant dust emissivity spectral index $\beta$, as
noted in Planck Collaboration XI (2014).  Many IVCs also appear in these 
difference plots (see Section 3.3).  
 }
\end{figure}
 


\begin{figure}[ht]
\includegraphics[angle=0,scale=1.1] {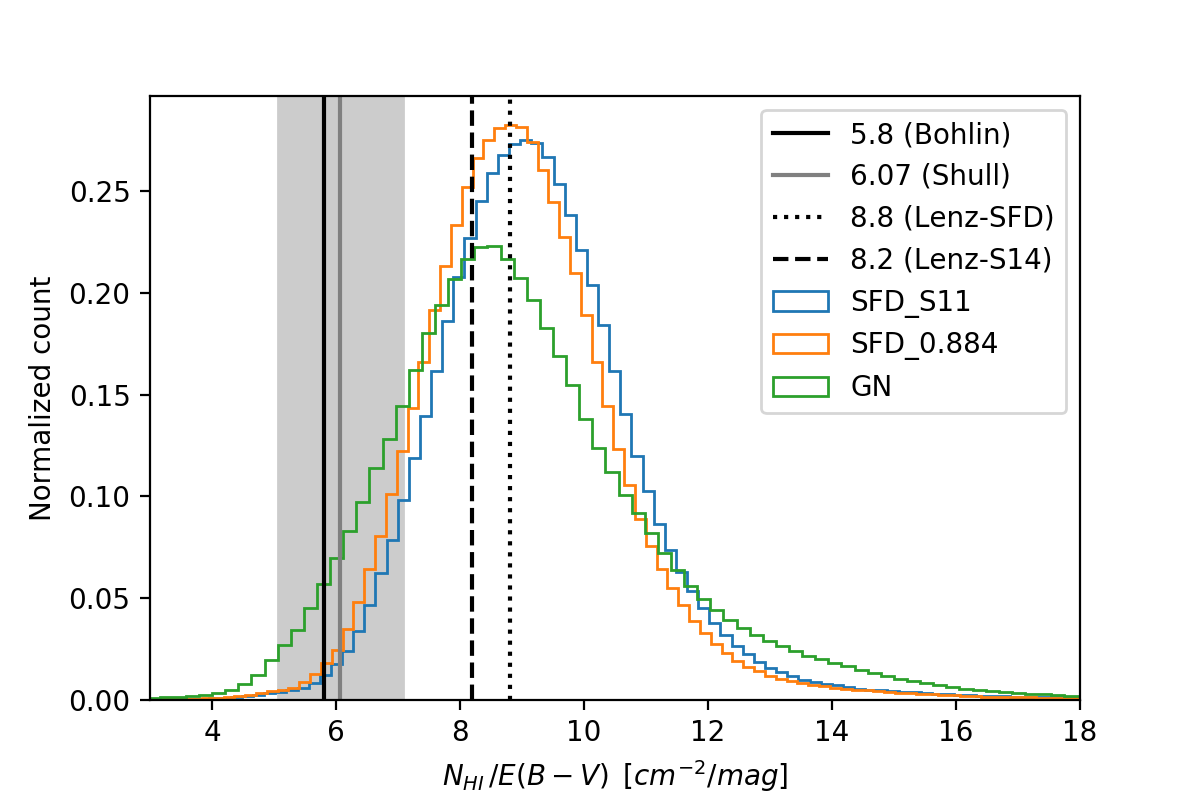}
\caption{Comparison of distributions of gas-to-dust ratios over the high-latitude sky
for three choices of reddening map  (SFD98, SF11, Planck-GN).  Here $N_{\rm HI}$ 
comes from HI4PI survey for gas at $|V_{\rm LSR}| \leq 90$~km~s$^{-1}$.  Reddening 
maps are smoothed to the HI4PI resolution ($16'$).  In the label box, SFD--S11 refers 
to the standard 0.86 recalibration, and SFD--0.884 refers to recalibration with the 
Johnson bandpasses, consistent with Planck-GN analysis. Each histogram is normalized
 to unit area.  Vertical lines show mean ratios determined in previous papers:  
 5.8 (Bohlin \etal\ 1978) and 6.07 (Shull \etal\ 2021) toward stars in the Galactic disk, 
and 8.8 and 8.2 quoted in Lenz \etal\ (2017) at high-latitude. Grey wash shows the 
($1\sigma$) variance in the Shull \etal\ (2021) survey of 129 stars within 5~kpc.  
 }
\end{figure}
 


\begin{figure}[ht]
\includegraphics[angle=0,scale=1.1]{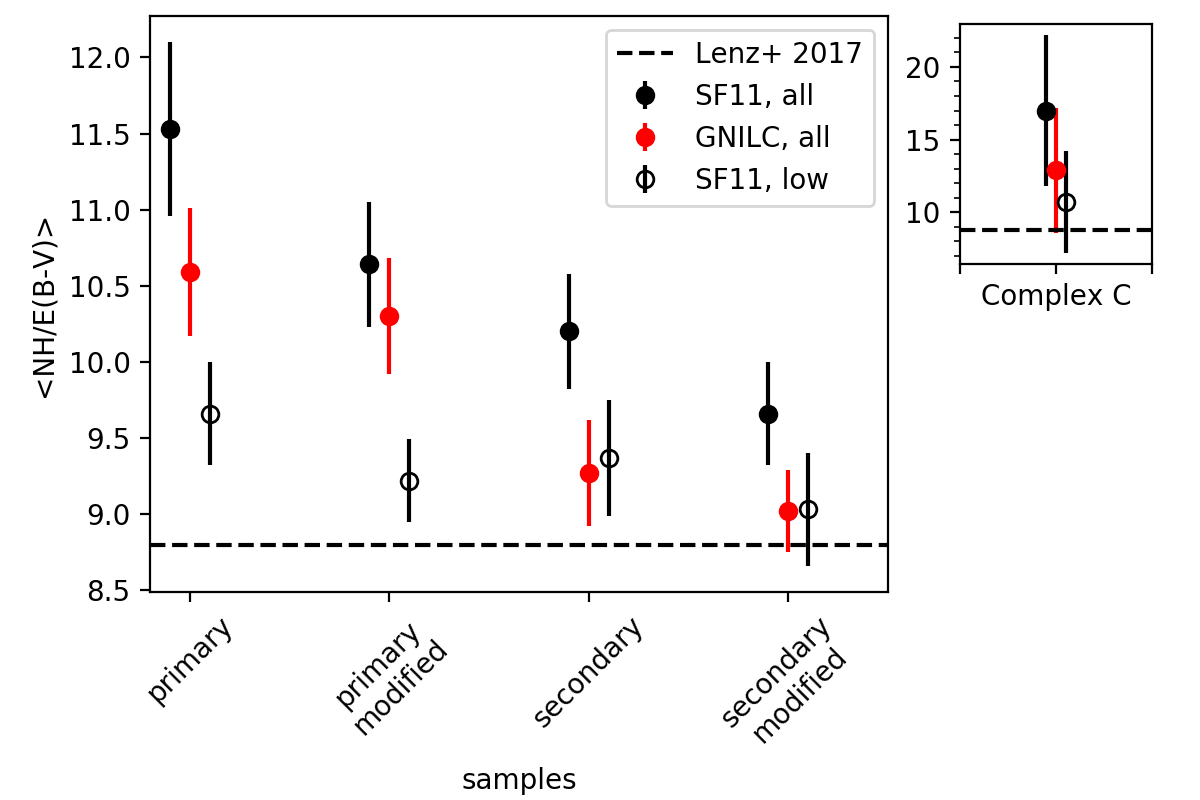}
\caption{Summary of mean ratios \Ratio\ from our primary sample (group 1, 55 AGN)
with both \HI\  and \Htwo\ measurements, and secondary sample (group 2, 39 AGN) 
with only \HI.  Modified samples exclude sight lines with uncertain reddening 
$E(B-V) < 0.01$ (see Table 6).  Ratios are shown for two reddening maps (SF11 and 
Planck-GNILC) and for gas at low velocity and all velocities (including HVCs and IVCs).  
Nine sight lines through HVC Complex~C with elevated ratios (low dust content) 
are a subset of the primary sample. The horizontal dotted line shows the mean 
ratio of $N_{\rm HI}/E(B-V) = 8.8$ at high latitude found by Lenz \etal\ (2017).  
 }
\end{figure}
 

\end{document}